\documentclass[acmtog,authorversion,timestamp]{acmart}

\acmSubmissionID{0076}
\setcopyright{none}

\usepackage{amsfonts}
\usepackage{color}
\definecolor{green}{rgb}{0, 0.5, 0}
\definecolor{orange}{rgb}{0.8, 0.6, 0.2}
\definecolor{red}{rgb}{1.0, 0.0, 0.0}
\definecolor{teal}{rgb}{0.0, 0.4, 0.4}
\definecolor{purple}{rgb}{0.65,0,0.65}
\definecolor{saffron}{rgb}{0.95,0.75,0.2}
\definecolor{turquoise}{rgb}{0.0,0.5,0.5}
\definecolor{black}{rgb}{0.0, 0.0, 0.0}

\newcommand{\rz}[1]{{\color{black}#1}}

\usepackage[normalem]{ulem}

\usepackage{amsopn}
\usepackage{mathtools}

\usepackage{amsmath, amsthm, amsfonts, amssymb}
\usepackage{graphicx} 
\usepackage{textcomp}
\usepackage{xfrac}

\usepackage{overpic}
\usepackage{graphicx}
\usepackage{float}
\usepackage{multirow}

\newcommand{\rmf}{\rm\bf}

\graphicspath{{figures/img/}}

\setcitestyle{square}

\begin{document}

\title{Semi-Supervised Co-Analysis of 3D Shape Styles from Projected Lines}

\author{Fenggen Yu}
\affiliation{%
	\institution{Nanjing University}}
\author{Yan Zhang}
\affiliation{%
	\institution{Nanjing University}}
\author{Kai Xu}
\affiliation{%
	\institution{National University of Defense Technology}}
\author{Ali Mahdavi-Amiri}
\affiliation{%
	\institution{Simon Fraser University}}
\author{Hao Zhang}
\affiliation{%
	\institution{Simon Fraser University}}

\renewcommand\shortauthors{Yu, F. et al.}

\begin{abstract}
\rz{We present a semi-supervised co-analysis method for learning 3D shape styles from
projective line drawings, achieving style patch localization with only weak supervision.}
%
Given a collection of 3D shapes spanning multiple object categories and styles, we initially perform
projective style co-analysis over projective line drawings of each 3D shape and then backproject
the learned style features onto the 3D shapes. Our core analysis pipeline
starts with mid-level patch sampling and pre-selection of candidate style patches. Projective
features are then encoded via patch convolution.
Multi-view feature integration and style
clustering are carried out under the framework of partially shared latent factor (PSLF) learning,
a multi-view feature learning scheme.
PSLF achieves effective multi-view feature fusion via distilling and exploiting
the consistent and complementary information from multiple views,
meanwhile selects style patches from the candidates.
Our style analysis approach supports both unsupervised and semi-supervised analysis.
For the latter, our method accepts both user-specified shape labels and
style-ranked triplets as clustering constraints.
We particularly demonstrate the effectiveness of our method for style analysis and patch localization and clarify improvements over state-of-the-art methods.
\end{abstract}



\maketitle


\section{Introduction}
\label{sec:intro}

\begin{figure}[htb]
  \centering

  \includegraphics[width=.99\linewidth]{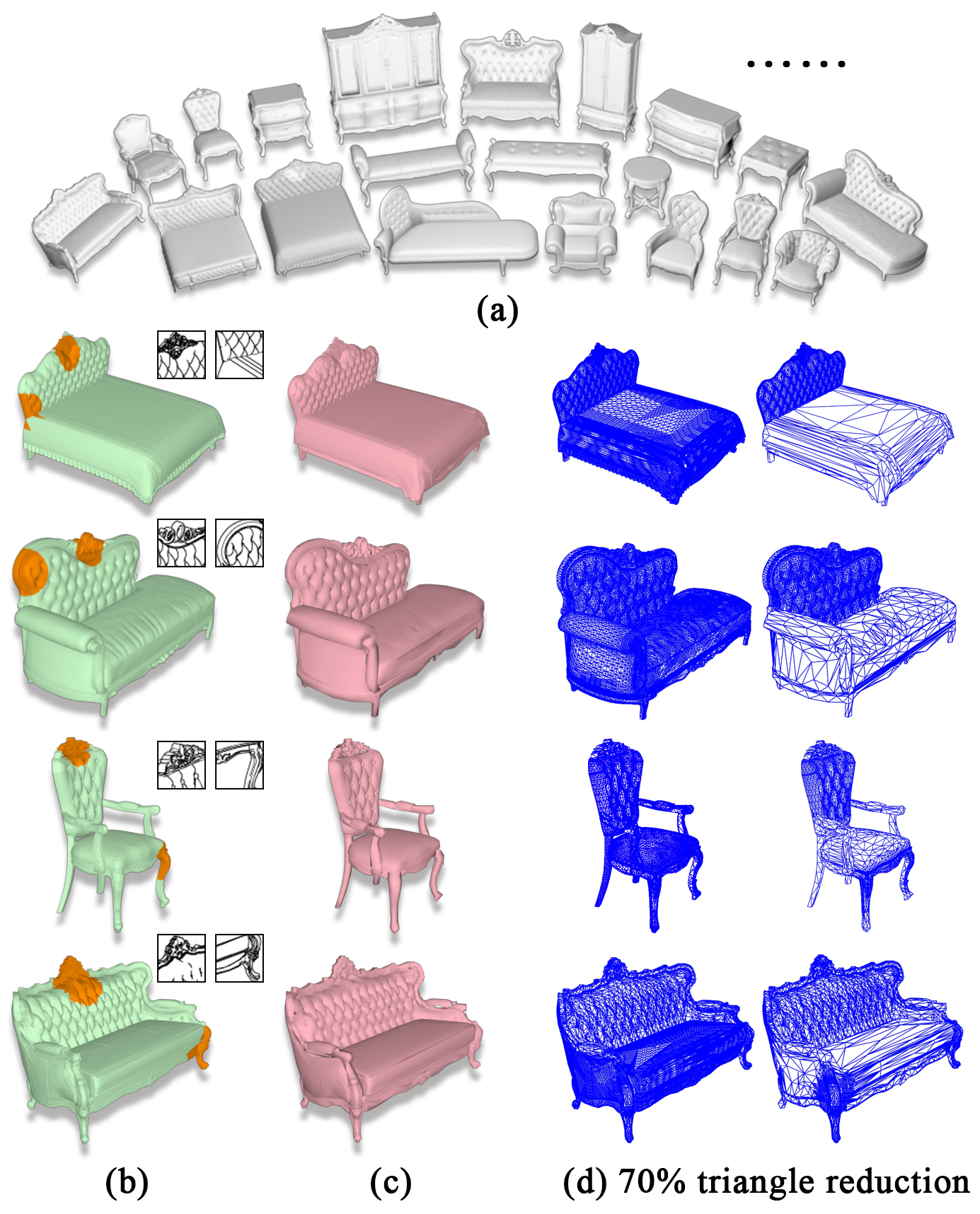}

  \caption{\label{fig:teaser}
   Given a heterogeneous shape collection (a), our method performs semi-supervised style co-analysis,
   over projected line drawings (see insets in left column), to spatially located style patches (b)
   and cluster the shapes based on their styles --- all the four shapes in color belong to the same
   cluster. Spatial localization of style patches enables applications such as style-preserving mesh
   simplification (c). 
   Triangle distributions before and after simplification are shown in (d).}
\end{figure}

Styles are generally regarded as distinctive and recognizable forms which permit the
grouping of entities containing these forms into related categories~\cite{wiki:vis_style}.
It follows that stylistic forms that serve to characterize a common style tend to share
strong similarities, while between different style categories, these forms often exhibit
clear distinctions. As a result, style analysis is best conducted in the context of a
{\em set\/} of entities and naturally lends itself as a {\em clustering\/} problem. For 2D
or 3D shapes, the shape styles are typically perceived by humans as apparent geometric
features or patterns; see Figure~\ref{fig:teaser} (left). The ability to extract such style features
allows them to be compared, altered, or preserved. 

\begin{figure*}[tb]
  \centering
  \includegraphics[width=.99\linewidth]{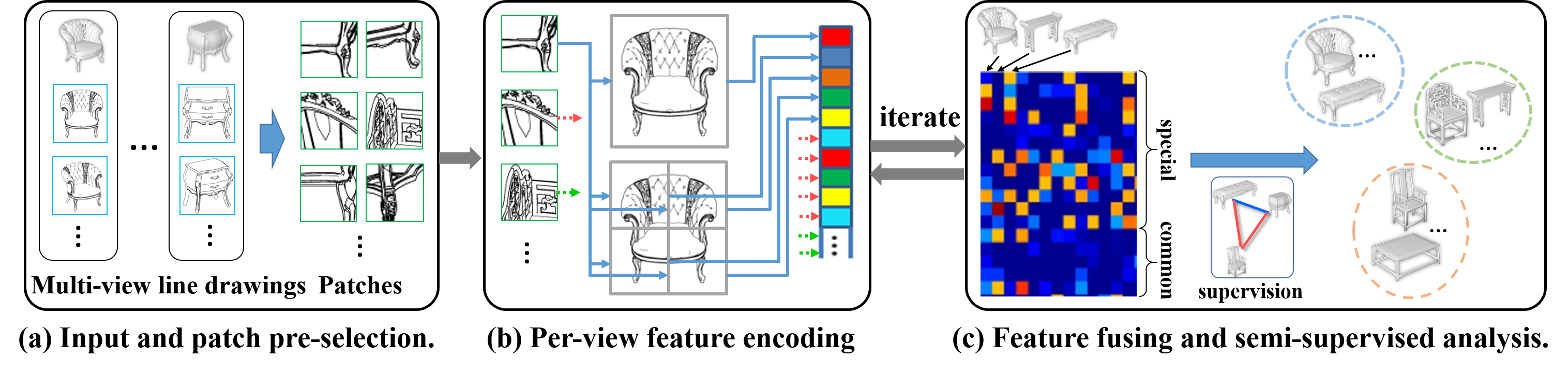}
  \caption{
Our style co-analysis algorithm contains three stages:
(a) Patch sampling and pre-selection.
(b) View feature encoding based on patch convolution, and
(c) Multi-view feature integration using partially shared latent factor (PSLF) learning.
The PSLF performs unsupervised or semi-supervised style clustering and patch filtering in an
interleaving fashion.}
\label{fig:overview}
\end{figure*}


Clustering analysis has been performed in earlier works on shape styles. 
However, the studied styles were either pre-determined~\cite{xu2010style} or characterized by hand-crafted
rules~\cite{li2013curve}. Most recent attempts have been on \rz{supervised\/} learning of style
similarity~\cite{garces2014,liu2015style,lun2015elements} via crowd-sourcing to collect user-specified style rankings
and then performing metric learning rather than style clustering. But these works do not spatially locate the stylistic
features or patches over the analyzed shapes. \rz{In a most recent work, also based on
supervised learning, Hu et al.~\shortcite{Style17} learn to spatially locate style-defining elements or patches over a set of 3D
shapes, where an expert-specified style clustering is given over the shape collection.\/}


\rz{
In this paper, we are interested in exploring a ``middle ground'', via
{\em semi-supervised\/} learning with {\em weak supervision\/}, for generic style analysis of 3D shapes.
Semi-supervised learning is attractive as it can take advantage of strong techniques for
unsupervised clustering and discriminative analyses without the need to collect large amounts of user data.
After all, style analysis is essentially a grouping problem, while style patch extraction is a
discriminative feature selection problem. That said, with semi-supervised learning, human is not
out of the loop. The learning process naturally incorporates user feedback to reflect the subjective
nature of style perception, while keeping such feedback to a minimum.\/}

\rz{
Specifically, we introduce a semi-supervised {\em co-analysis\/} method which simultaneously achieves
style clustering and style patch localization, 
with only weak supervision over a {\em heterogeneous\/} collection of 3D shapes spanning multiple
object categories and styles. Unlike Hu et al.~\shortcite{Style17}, the input collection is not clustered
by experts, our method supports both unsupervised and weakly-supervised analysis with minimal style
annotation. In terms of shape styles, like all previous works~\cite{liu2015style,lun2015elements,Style17}, our analysis
also focuses on {\em element-level\/} styles of 3D shapes~\cite{lun2015elements} which are {\em decorative\/} in nature.\/}
Such styles include those that are perceivable as patterns along shape contours or over
shape surfaces. They are ubiquitous in man-made shapes including furniture, buildings,
automobiles, kitchen utensils, and many engineering products and household items.

%
%

Our core analysis problem consists of a {\em clustering\/} of the input shape collection and
a {\em selection of style feature\/} patches from the shapes which accentuate the shape clustering.
While latent features are sufficient for style comparisons, to spatially locate shape styles,
one must eventually extract and discriminate between {\em spatially explicit\/}
or {\em visually apparent\/} shape features. Working with these features for style analysis is
also well motivated by the visual and perceptual nature of style recognition by humans:
styles are {\em seen\/} as visual patterns. In our work, we perform style analysis of 3D shapes
via {\em projective\/} analysis. Specifically, we project a 3D shape over different views and work
with projective {\em line drawings\/} for our style analysis and the extraction of shape styles.

Even though line drawings do not exist in the ``real world'', they are believed to possess
deep similarities to other more detailed and explicit visual representations as well as real
scenes they depict~\cite{sayim2011}. In addition, they are remarkably efficient
at conveying shape and meaning while reducing visual clutter~\cite{szymon2008}. For our
purpose, line drawings are well-suited to depict decorative style patches of 3D shapes.
%
%
On the technical front, projective analysis puts many effective learning methods designed for
image data, e.g.,~\cite{bansal2015mid,gong2014multi,Li2015Mid}, at our disposal. Furthermore, it
works more robustly with different 3D geometry representations and various shape imperfections
including noise, incompleteness of shapes, and non-manifold geometries~\cite{wang2013projective}.
We also emphasize that our projected lines include \emph{geometric feature lines} which cannot
be captured by rendered images, although the neuron activations in CNN could be used to extract
lines and corners from rendered images. Experiments show that line drawing outperforms color rendering,
even when both using CNN for feature extraction.


Given a heterogeneous collection of 3D shapes spanning multiple object categories and
styles, our method performs style co-analysis over projective line drawings of each 3D shape and
backproject the learned style features onto the 3D shapes at the end. As shown in
Figure~\ref{fig:overview}, our core analysis pipeline consists of three stages: 1) mid-level patch sampling
and pre-selection of candidate style patches; 2) view feature encoding based on patch convolution; and 3)
multi-view feature fusion and style clustering under the framework of {\em partially shared latent factor
learning\/} or PSLF~\cite{liu2015partially}, which selects the final style patches from the candidates.

PSLF fits well with projective analysis, as it is designed for multi-view feature analysis and learning.
It deciphers the consistent and complementary information from the features of multiple
views and integrate them in a better informed way.
We show that such an advanced feature fusion scheme performs much better than the max-pooling used in
multi-view CNN~\cite{su2015multi}.
Furthermore, PSLF discovers shape styles by clustering shapes and selecting the most discriminative mid-level patches
which accentuate the clustering; this is consistent with how styles are typically characterized~\cite{wiki:vis_style}.
To support both unsupervised and semi-supervised style analysis, we develop a {\em constrained\/}
formulation of PSLF which accepts both user-specified shape style labels~\cite{Style17} and style-ranked triplets as
classical metric learning~\cite{garces2014,lun2015elements,liu2015style}.

Figure~\ref{fig:teaser} shows a sample result, where our style co-analysis was performed on 400 mixed furniture pieces (top row).
Four shapes deemed to belong to the same style cluster (not all shapes in the cluster appear in the figure) are shown with their
style patches highlighted, both on the shape and also in line drawings (see insets).

Our work makes the following main contributions:
\begin{itemize}
\item \rz{To the best of our knowledge, our method represents the first semi- or weakly-supervised
co-analysis of 3D shape styles, leading to style clustering and style localization.\/}
\item Our method supports both unsupervised and semi-supervised style analysis,
combining local feature learning and global discriminative style extraction.
\item \rz{Our analysis focuses exclusively on projective features from line drawings, while previous
works employed much richer feature sets~\cite{lun2015elements,liu2015style,Style17}. Yet, as we shall
demonstrate, we can obtain clear improvements over all of these methods, owing in part to our
multi-view style analysis with feature fusion.\/}
\item We can spatially locate visually apparent stylistic shape elements or patches without any direct user involvement to
manually mark any style patches over 3D shapes. Our semi-supervised analysis takes the same types of
user input as classical metric learning.
\end{itemize}

We demonstrate the effectiveness of our method for style analysis and patch localization, in particular, clear
improvements over state-of-the-art supervised methods~\cite{lun2015elements,liu2015style,Style17}. We also
develop several applications that can take advantage of the detected styles. For example, with the style
patches spatially located, we can perform style-preserving mesh simplification, as shown
in Figure~\ref{fig:teaser}. Triangle distributions before and after simplification, shown in
Figure~\ref{fig:teaser}(d), clearly exhibit that triangle reduction happens mostly over non-style regions. 
\section{Related work}
\label{sec:related}

In this section, we cover and discuss works related to shape style analysis and our approach
for style clustering and style patch extraction on 3D surfaces via machine learning.
We describe how our method is different from these existing approaches.

\vspace{-5pt}

\paragraph{Co-analysis.}
Our approach falls in the realm of co-analysis techniques~\cite{mitra2013,zhu2018scores}, most of
which have been designed to work with homogeneous shape collections. Our method can
work with a heterogeneous shape collection owing to its localized feature encoding and
analysis of decorative styles. Semi-supervised learning has been mainly employed to
solve labeling problems such as shape segmentation~\cite{wang2013projective} and
classification~\cite{huang2013fine,zhao2018triangle}. In our work, we rely on unsupervised and semi-supervised
PSLF learning to extract spatial style patches.

\vspace{-5pt}

\paragraph{Projective shape analysis.}
Analyzing 3D shapes through 2D projections has been a common practice with successful
applications such as shape classification~\cite{su2015multi}, retrieval ~\cite{chen2003visual},
and segmentation ~\cite{wang2013projective}, to name a few. Main benefits of the projective
approach include robustness with imperfect 3D representations and exploitation of image-based
learning, especially deep learning techniques. Our work offers a new application, namely analysis
of shape styles, by utilizing another useful property of 2D projections, i.e, their ability to reveal
shape styles visually in the form of line drawings.

\vspace{-5pt}

\paragraph{Multi-view learning.}
In multi-view learning, a concept is learned from data represented in multiple forms or
views, e.g.~\cite{chaudhuri2009multi,liu2015partially}. The goal is
to discover \emph{consistent} and \emph{complementary} information among
multiple views of the data, with both types of information supporting the
concept. Specifically, consistent information should be shared across most views
in identifying the concept, while complementary
information is something that is distinctively reflected from one or few views and
complementary to other such information.
In our work, the concept to learn is shape styles and the multiple views are provided
by the multi-projection line drawings of the 3D shapes. Decorative shape styles may be shared
consistently across multiple views, e.g., stylistic details over the surface of a sofa. They can be also exclusive
to one or few views to complement other style features such as an emblem on top of a bed's headboard. As a result, our style analysis problem is well-suited for a multi-view learning approach.


\vspace{-5pt}

\paragraph{PSLF}
Partially shared latent factor (PSLF) learning is a multi-view learning method~\cite{liu2015partially}
which performs joint analysis over a set of data to extract {\em both\/} the consistent and
complementary information among multiple data views. PSLF is essentially a dimensionality reduction
technique that is realized through a non-negative matrix factorization (NMF). Our method adopts
PSLF to learn style features and their spatial locations via projective co-analysis. We also adjust
the original objective function of PSLF to enable semi-supervised learning that accepts both
user-specified shape labels and style-ranked triplets as classical metric learning.


\vspace{-5pt}

\paragraph{Mid-level patches.}
Mid-level image patches, e.g., object parts or salient regions, are neither too
local nor too global and they have been effective for tasks such as object
detection~\cite{bansal2015mid}, indoor scene classification~\cite{doersch2013mid}, and unsupervised
visual discovery~\cite{raptis2012,singh2012unsupervised}. While unsupervised approaches
typically perform the analysis purely at the patch level~\cite{singh2012unsupervised}, weakly supervised
approaches such as those presented by~\cite{doersch2013mid,bansal2015mid}, typically
detect mid-level patches or compute patch clusterings to attain maximal adherence to the image or object labels.

Most closely related to our method is the work by Lee et al.~\shortcite{lee2013style}, which aims to discover
mid-level patches that are the characteristic of historic and geographic styles of objects in images. They start
with a generic visual element detector serving a similar role as our pre-selection of
patches, and then rely on image labels of date and location to train a regression model to
identify style-aware mid-level patches. In contrast, we rely on PSLF to perform shape clustering
and style patch selection in an interleaving manner. Our approach can be supervised or semi-supervised,
and for the latter, both style labels and user-specified style rankings are accommodated.

\vspace{-5pt}

\paragraph{Convolutional activation features.}
Convolutional neural nets (CNNs) have been extensively employed for various
feature learning tasks recently, e.g.,~\cite{donahue2013decaf,razavian2014cnn,zeiler2014visualizing}. 
Training a full CNN for feature learning is expensive in terms of data labeling and computation.
In our work, we explore the limit of unsupervised and semi-supervised feature learning for shape
style analysis requiring minimal data annotations.

Deep convolutional activation features, including those for mid-level
visual elements, have been employed as descriptors for generic visual recognition~\cite{bansal2015mid,gong2014multi,Li2015Mid}.
We use the discriminatively detected mid-level patches as filters to perform feature encoding
based on a sliding-window convolutional operation, similar to~\cite{bansal2015mid}.
Although these content features are discriminative for object characterization,
it is not yet clear whether they would attain the same level of success for style
recognition since style and content features do not always correlate with each other.
We use these features as per-view initial features and conduct multi-view feature fusion and selection via PSLF.

%


\vspace{-5pt}

\paragraph{Shape style analysis.}
Some earlier works on shape styles assume that the style is given. For example, Xu et al.
\shortcite{xu2010style} have worked exclusively with anisotropic part proportions. Some other
works perform style analysis on shapes which belong to the same semantic
category~\cite{huang2013fine,kalogerakis2012probabilistic}. More recent attempts
generally take the view that human style perception transcends shape content~\cite{lun2015elements}.
In an earlier work, Li et al.~\shortcite{li2013curve} have handcrafted several rules as an attempt to
characterize style features for 2D curves. Another line of works focus on analogy-based style
transfer, e.g., Ma et al.~\shortcite{ma2014analogy}, where the goal is to determine what editing
operations on a query shape $A'$ mimic the style change which would transfer a given shape
$A$ to a given shape $B$. Lun et al.~\shortcite{lun2016} have shown that the same style transfer
framework can be extended with deep learning techniques.

\vspace{-5pt}

\paragraph{Supervised learning of style similarity.}
Recent works on shape style analysis combine crowd-sourcing and metric learning
to learn a generic style similarity. Most notably, Lun et al.~\shortcite{lun2015elements} and
Liu et al.~\shortcite{liu2015style} both work with heterogeneous 3D shape collections and
rely on crowd-sourced style ranking triplets to learn a style metric. Most recently, Lim et al.
\shortcite{lim2016} added deep learning to this framework. The key difference between our 
method and these works is that we learn to spatially locate style patches and they do not. 
\rz{Also, there are differences in what is learned and for what target applications. Our
work learns what makes a piece of furniture Chinese/Country and a building 
Gothic/Greek and where the style regions are. The core problems we face are 
style classification/clustering and style patch extraction. In contrast, Lun et al.~\shortcite{lun2015elements} and
Liu et al.~\shortcite{liu2015style} focused on style-driven shape retrieval, trying to learn a 
specialized shape similarity. Yet, our method can be customized to accomplish tasks Lun et al.~\shortcite{lun2015elements} and
Liu et al.~\shortcite{liu2015style} were designed to do, making it more general.\/}

On the technical front, several other
differences exist: 1) our method can be both unsupervised and semi-supervised; 2) our method
employs projective analysis and relies on different features; 3) our method adapts PSLF clustering
to help us select and locate style feature patches, while both Liu et al.~\shortcite{liu2015style}
and Lun et al.~\shortcite{lun2015elements} adapt metric learning to learn a global style metric.

\vspace{-5pt}

\rz{
\paragraph{Supervised style localization.}
The work by Hu et al.~\shortcite{Style17} also learns to locate style patches, but it takes as input a 
collection of 3D shapes with expert-provided style clustering. In contrast, the input to our work is only such 
a shape collection, {\em without any style clustering\/}. In our semi-supervised version, the user can provide 
style labels or style ranking triplets, but only over a very small percentage of the data. Therefore, their work
is exclusively on feature selection based on a given style clustering while we need to solve both clustering
and feature extraction simultaneously. 

Another difference, a subtle one, is that Hu et al.~\shortcite{Style17} aim to locate {\em style-defining\/}
patches, i.e., all patches which together define a particular style, while our analysis seeks to find
{\em style-discriminative\/} patches, i.e., those which can distinguish a style from the others. In their work,
style-defining patches are simply a combination of style-discriminative patches.
Technically, the feature learning schemes of the two methods are different and they 
operate on different features: we work with projective line drawings and Hu et al.~\shortcite{Style17} 
employed similar features as Lun et al.~\shortcite{lun2015elements}.
}
\section{Overview}
\label{sec:overview}

Given a heterogeneous collection of 3D shapes in several styles, we perform projective style analysis based
on multi-view line drawings of each 3D shape. Our method contains three stages:
patch sampling and pre-selection, view feature encoding based on patch convolution, and multi-view feature
integration with partially shared latent factor (PSLF) learning.
The PSLF interleaves style clustering and patch filtering in an unsupervised or semi-supervised fashion.
Figure~\ref{fig:overview} illustrates an overview of our method.

\begin{figure}[tb]
  \centering

  \includegraphics[width=.99\linewidth]{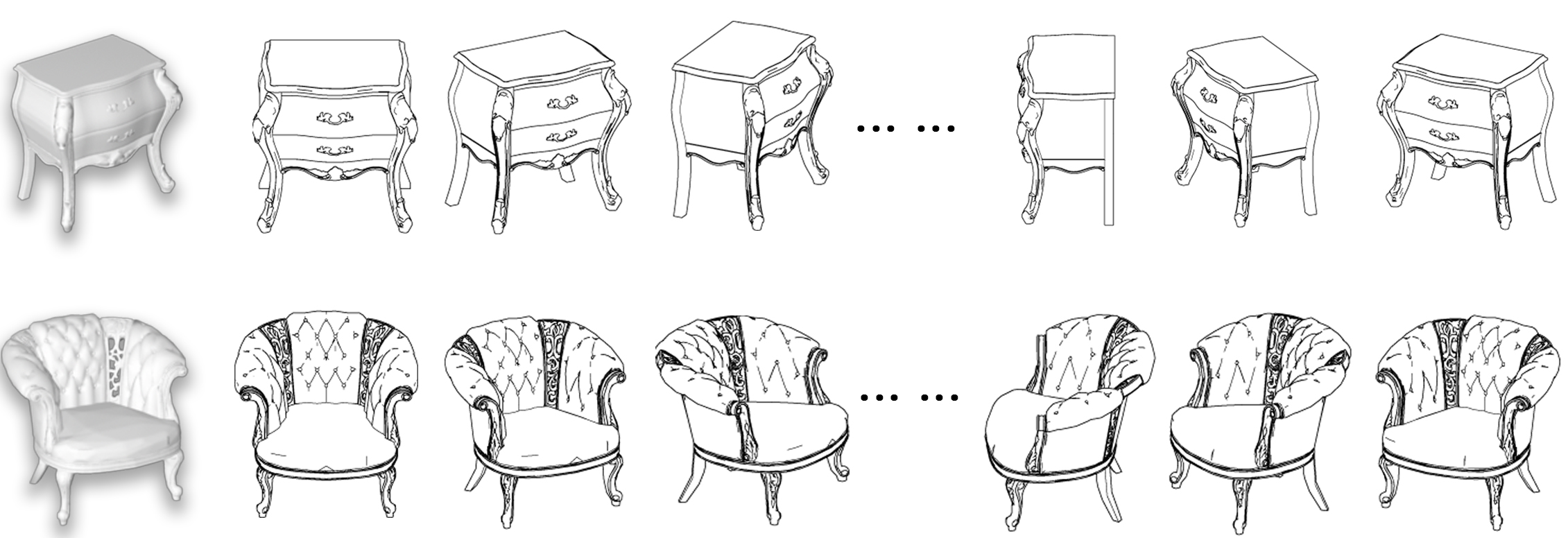}

  \caption{\label{fig:linedrawing}
           Some multi-view line drawings from 3D shapes.}
\end{figure}

\vspace{-5pt}
\paragraph{Multi-view line drawing.}
For each 3D shape, we render it from the views of $12$ virtual cameras located circularly around the shape
in every $30$ degrees. These cameras are elevated $30$ degrees from the ground, pointing towards the centroid
of the shape.
For each view, we extract both suggestive contours~\cite{decarlo2003suggestive} and dihedral angle based feature lines~\cite{gal2009iwires},
leading to an image of $200\times200$ size.
While the former captures contours and creases of a smooth manifold, the latter is especially useful
for extracting sharp feature lines from man-made shapes which can be potentially non-manifold.
See Figure~\ref{fig:linedrawing} for a few examples of multiple-view line drawings.

\vspace{-5pt}

\paragraph{Patch sampling and pre-selection.}
We select a set of representative mid-level patches from all projected line drawing images,
that are used as the convolutional kernels in the feature encoding of the projections.
Specifically, we first randomly sample a set of points on each shape.
For each sample point and each view in which this point is visible, a patch is generated as the window centered at the projection of the point.
We then perform $k$-means clustering to extract a set of representatives as the cluster centers (Figure~\ref{fig:overview}(a)).

\vspace{-5pt}

\paragraph{Per-view feature encoding.}
In the second step, a feature map is extracted for each line drawing image through convolving it using the pre-selected patches as kernels.
Convolutional feature encoding is known to be shift-invariant, since a convolution kernel may get activated at an arbitrary position in an image.
This trait fits well to our problem since local style patches may appear in multiple spatial locations.
To extract multi-scale features, we also perform convolution for sub-images (Figure~\ref{fig:overview}(b)).
The final feature is a concatenation of the per-region feature after pooling, similar to~\cite{bansal2015mid}.

\vspace{-5pt}

\paragraph{Multi-view feature integration.}
The core step of our algorithm is to fuse the features extracted from different views while clustering the shapes based on the fused features. This leads to a multi-view feature representation for each shape. We adopt the partially shared latent factor (PSLF) learning~\cite{liu2015partially} to implicitly separate the input multi-view features into parts which are shared by multiple views and those which are distinct to a specific view. The final multi-view feature is compact and comprehensive, encoding both shared and distinct information in different views.

\vspace{-5pt}

\paragraph{Unsupervised and semi-supervised style analysis.}
Based on the clustering result, we re-select the representative mid-level patches to
learn more and more discriminative ones with respect to the evolving style clusters.
This will in turn update the feature encoding in the next iteration.
Such cluster-and-select process iterates until the clusters and patches become stable. Our process can be performed unsupervised to cluster models.
To impart human knowledge about shape styles into the analysis, we realize semi-supervised style
clustering within the PSLF framework, achieving both meaningful style clustering and informative feature learning.
Specifically, we present two semi-supervised clustering methods, accepting either user-prescribed style labels
on a small portion of the shape collection or triplets of shapes indicating their style similarity (Figure~\ref{fig:overview}(c)).
Finally, we backproject the learned discriminative patches from projective space to surfaces of
the input 3D shapes to extract and visualize the style patches over these 3D shapes.

\section{Semi-supervised projective style co-analysis}
\label{sec:method}

In this section, we describe our semi-supervised projective style co-analysis method in detail.

\subsection{Patch sampling and pre-selection}
\label{subsec:patch}


We first sample 2D patches from the multi-view rendered line drawings to bootstrap our style analysis.
To ensure a uniform coverage of a shape surface,
instead of sampling the patches in 2D projections, we sample 3D points
on the shape surface and then use the 3D points as seeds to generate 2D patches through projection.
3D sampling also facilitates the back-projection of 2D patches into 3D for locating style patches.
In practice, we sample $30$ seed points on a 3D surface, project them onto the 2D views in which these points are visible.
We then, for each projected 2D point, we extract a 2D square patch centered at this point.
\rz{For a $200\times200$ image, we extract about $30$ patches, where the patch size is chosen experimentally; see discussion in Section~\ref{sec:results}.}



\begin{figure}[tb]
  \centering

  \includegraphics[width=.99\linewidth]{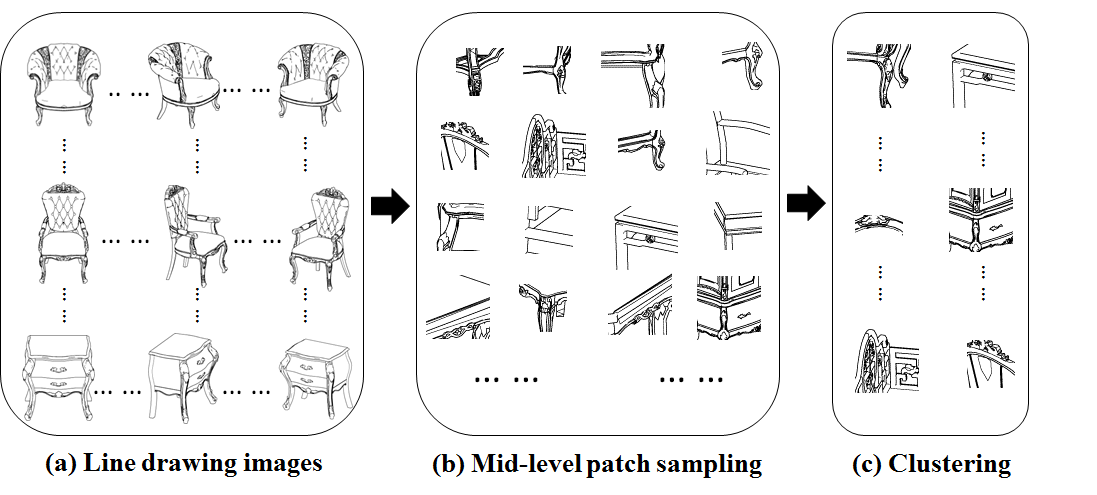}

  \caption{\label{fig:PatchSampling}
           Patch sampling and pre-selection.
           (a) The input line drawings.
           (b) The initial patch sampling.
           (c) Mid-level patches obtained by k-means clustering.}
\end{figure}

To select a set of representative patches for each view,
we perform $k$-means clustering over all patches in that view in HOG feature space~\cite{dalal2005histograms}.
The cluster centers are selected as the representative patches.
In practice, we extract $50$ representative patches for each view.
Figure~\ref{fig:PatchSampling} demonstrates the sampled mid-level patches as well as the selected representatives.

\subsection{Per-view feature encoding}
Inspired by the work of Bansal et al.~\shortcite{bansal2015mid}, we perform convolutional feature encoding for each line drawing image
using the representative mid-level patches as convolution filters.
While these features are extracted by directly applying convolution operations over input images,
without the need of training a deep neural network,
they were shown to perform comparably against the deep CNN features on 2D object detection task.
The main rationale behind this is that the characteristic mid-level patches,
analyzed from a relevant image collection~\cite{bansal2015mid},
encompass informative visual cues for an object class.
Our iterative patch selection is conducted over relevant image collections obtained by clustering.

\begin{figure}[b]
  \centering
  \includegraphics[width=.99\linewidth]{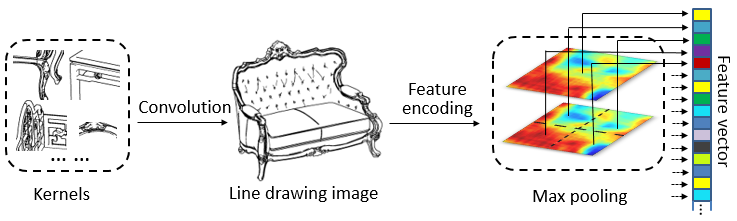}
  \caption{\label{fig:FeatureRepresentation}
           Feature representation.
           For each line drawing, after performing convolution via each convolution kernel followed by max-pooling, we obtain a new feature vector.}
\end{figure}

We take one mid-level patch as a convolution filter and use it to convolve the input image in a sliding-window fashion.
Such a convolution operation is conducted in HOG feature space: both the input image and mid-level patches are represented by HOG feature maps.
To compensate the global feature encoding of full image convolution,
we also perform the above process over the sub-image obtained by dividing the original image into four parts.
We then perform max-pooling over the convolution activations over the spatial pyramid of two levels of resolution (Figure~\ref{fig:FeatureRepresentation}).
Consequently, each convolution filter produces a $5$-dimensional feature vector.
The ultimate feature for a line drawing image is constructed by concatenating the feature vectors of all convolution filters,
leading to a $5K$-dimensional feature vector for $K$ mid-level patches.

\subsection{Multi-view feature fusion}
\label{subsec:fusion}

Having a feature vector for each view of a given shape, we perform multi-view feature fusing, 
to extract a new feature for the shape. This feature instils the information from jointly 
analyzing a set of shapes. To achieve this, we adopt the partially shared latent factor (PSLF) framework~\cite{liu2015partially}, which is a clustering method coupled with multi-view 
feature integration. Note that the ``views'' in the original work generally refer to different 
aspects, attributes, or observations of the data in question. In our case, the views are given 
by line drawings in multiple projections of 3D shapes.

\begin{figure}[tb]
  \centering
  \includegraphics[width=.99\linewidth]{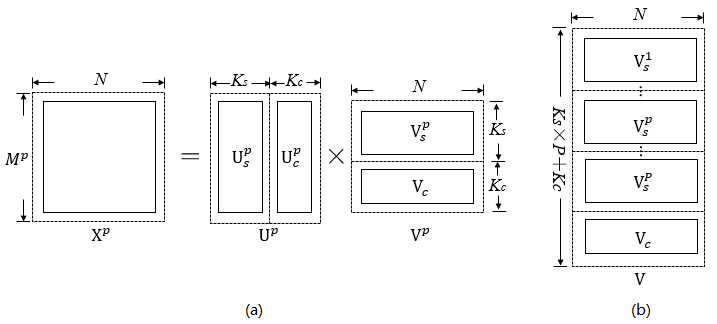}
  \caption{\label{fig:Matrix}
           Matrix factorization in the $p$-view and required partially shared latent factor matrix.
           (a) Input matrix ${\rmf{X}}^p$ represents feature vector matrix of a shape collection in $p$-view.
           (b) Partially shared latent factor matrix ${\rmf{V}}$.}
\end{figure}

PSLF employs non-negative matrix factorization (NMF)~\cite{lee1999learning} to learn a compact yet comprehensive partially shared latent
representation. Given a collection of $N$ shapes with line drawings in $P$ views,
the learning objective of PSLF is:
\begin{align}\label{eq:pslf}
  & {\rm{min.}}\quad\sum_{p = 1}^P {{\pi ^p}||{{\rmf{X}}^p} - {{\rmf{U}}^p}{{\rmf{V}}^p}||_F^2 + \lambda ||\varPi ||_2^2}, \\
  & \text{s.t.}\quad{{\rmf{U}}^1},\ldots,{{\rmf{U}}^p}, {{\rmf{V}}^1},\ldots,{{\rmf{V}}^p},\varPi  \ge 0,\sum_{p = 1}^P {{\pi ^p} = 1}, \nonumber
\end{align}
where the \emph{input feature} matrix ${\rmf{X}}^p \in{\mathbb{R}^{{M} \times N}}$ contains the per-view feature of all $N$ shapes, for
the $p$-th view, with each column corresponding to one shape.
$M$ is the length of the feature vector of a given shape.
${\rmf{U}}^p \in{\mathbb{R}^{{M} \times K}}$ is the basis matrix of view $p$, while
${\rmf{V}}^p \in{\mathbb{R}^{K \times N}}$ is the matrix of $K$ latent factors, $K \ll M$, or the \emph{fused feature} matrix.
$\varPi=(\pi^1,\pi^2,\ldots,\pi^p)$ is the weights for different views.
$\lambda$ controls the smoothness of $\varPi$. A large value for $\lambda$ leads to smoother view weights.
Essentially, PSLF learns the fused feature matrix $\rmf{V}$ and the basis matrix $\rmf{U}$, while tuning the weights of different views,
all in an unsupervised manner, by minimizing the reconstruction error with respect to the input features.
The projection of the fused features over the basis leads to a clustering of input features.
The PSLF factorization for a set of input features in one view is illustrated in Figure~\ref{fig:Matrix}(a).

PSLF assumes that only parts of the latent factors are shared across all views and the other ones are separately embedded
in individual views. 
Thus, the factor matrix of view $p$ is separated into two parts: ${\rmf{V}}^p=\left[{\rmf{V}}_s^p,{\rmf{V}}_c\right]$, where ${\rmf{V}}_s^p$ represents the specific information extracted from view $p$ and ${\rmf{V}}_c$ the common shared by all views.
The basis matrix is also divided into two parts: ${\rmf{U}}^p=\left[{\rmf{U}}_s^p,{\rmf{U}}_c^p\right]$, with ${\rmf{U}}_s^p$ being the specific part corresponding to the shared latent factors and ${\rmf{U}}_c^p$ the common part.

An important parameter of PSLF is the proportion of common part: $\eta = (K_c/(K_s + K_c))$,
where $K_c$ and $K_s$ are respectively the dimensions of the common and specific latent factors and $K_s+K_c=K$ holds.
In this setting, when $\eta$ is larger, the role of consistency is more.

After performing the feature matrix factorization for all shapes being co-analyzed,
we obtain the partially shared latent factor matrix ${\rmf{V}} \in{\mathbb{R}^{({K_s} \times P + {K_c}) \times N}}$,
which contains the fused feature for each shape.
The matrix ${\rmf{V}}$ contains the unique feature part of each view ${\rmf{V}}_s^1,\ldots,{\rmf{V}}_s^p$
and the common part for all views ${\rmf{V}}_c$. (i.e., ${\rmf{V}} = \left[{\rmf{V}}_s^1;\ldots;{\rmf{V}}_s^p;{\rmf{V}}_c\right]$), see Figure~\ref{fig:Matrix}(b) for illustration.

An important feature of PSLF is that it is able to learn both consistent and complementary information from
the data views. When applied to style analysis, our experiments (see Section~\ref{sec:results}) confirm that both types
of information affect the learned shape styles, demonstrating the adaptability of PSLF to
our problem.

\subsection{Style clustering}
\label{subsec:stylecluster}


PSLF was originally proposed for multi-view, semi-supervised clustering and feature learning~\cite{liu2015partially}. To accommodate both user-provided style ranking triplets 
and style labels to constrain semi-supervised analysis using the learning framework 
provided by PSLF, we must modify the original PSLF formulation.


\vspace{-5pt}

\paragraph{Label-constrained style clustering.}
We modify the original objective function of PSLF to incorporate user-specified labels in constrained clustering.
Suppose ${\mathbf{V}}_l \in{\mathbb{R}^{({K_s} \times P + {K_c}) \times {N_l}}}$ is the feature vector matrix of $N_l$ number of shapes with labels.
Assume, w.l.o.g., the first $N_l$ columns of $\mathbf{V}$ correspond to the labeled shapes.
$\mathbf{Y} \in{\mathbb{R}^{C \times N_l}}$ is the corresponding label matrix, where
$C$ is the number of style categories. The objective we optimize for is:

\begin{equation}
\label{eq:pslf2}
{\min}\sum \limits_{p = 1}^P {\pi ^p}||{{\mathbf{X}}^p} - {{\mathbf{U}}^p}{{\mathbf{V}}^p}||_F^2 + \lambda ||\pi {\rm{||}}_2^2 +
\beta || {{\mathbf{V}}_l} - {\mathbf{WY}} ||_F^2 + \gamma ||\mathbf{W}||_{2,1}
\end{equation}
\[ \text{s.t.}\quad {{\mathbf{U}}^1},...,{{\mathbf{U}}^p},{{\mathbf{V}}^1},...,{{\mathbf{V}}^p},\varPi  \ge 0,\sum_{p = 1}^P {{\pi ^p} = 1} \]

where $\pi^p$, ${\mathbf{X}}^p$, ${\mathbf{U}}^p$, ${\mathbf{V}}^p$ and $\lambda$ are defined the same as before.
$\mathbf{W} \in{\mathbb{R}^{({K_s} \times P + {K_c}) \times C}}$ is the basis matrix obtained for labeled shapes.
$\beta > 0$ is a parameter for tuning the importance of user-specified labels.
$\gamma$ controls the weight of $\ell_{2,1}$ regularization term.

In this new optimization, $\beta || {{\mathbf{V}}_l} - {\mathbf{WY}} ||_F^2 + \gamma ||\mathbf{W}||_{2,1}$  is the semi-supervised term,
where the NMF of ${\mathbf{V}}_l$ produces the basis matrix $\mathbf{W}$ constrained with $\mathbf{Y}$.
Note that this optimization not only predicts cluster labels for unlabeled shapes, but it also updates the fused feature matrix ${\mathbf{V}}$.
Therefore, the output fused features in ${\mathbf{V}}$ have also incorporated the user constraints.

Since the basis $\mathbf{W}$ defines the cluster centers obtained from constraints,
the fused feature matrix of the $(N-N_l)$ unlabeled shapes, ${\mathbf{V}}_u \in{\mathbb{R}^{({K_s} \times P + {K_c}) \times (N-N_l)}}$
(so we have ${\mathbf{V}}=\left[{\mathbf{V}}_l,{\mathbf{V}}_u\right]$) can be defined by:
${\mathbf{V}}_u={\mathbf{WY}}_u$.
${\mathbf{Y}}_u \in{\mathbb{R}^{C \times (N-N_l)}}$ is the label prediction matrix for the unlabeled shapes.
These shapes can be assigned with the label corresponding to the largest probability.

\vspace{-5pt}

\paragraph{Triplet-constrained style clustering.}
\label{subsec:Tripletcluster}
Exact style labels are sometimes hard to perceive even by humans. It is relatively easier to provide similarity-based supervision,
e.g., shape A is style-wise closer to B than to C. This kind of user input has been extensively used in crowd sourcing~\cite{liu2015style,lun2015elements}.
To incorporate triplet-based constraints into our clustering, we decompose each triplet into two pair-wise constraints, i.e.,
\emph{must-link} and \emph{cannot-link} between a pair of data points, which is a standard form of constraints used by semi-supervised learning~\cite{chen2008non}.

Our method imposes such pair-wise constraints over the similarity matrix obtained by the unsupervised analysis of PSLF:
$\mathbf{A} = \mathbf{V}^T\mathbf{V}$, where $\mathbf{V}$ is the partially shared latent factor matrix discussed in Section~\ref{subsec:fusion}.
Specifically, we modify the similarity matrix as follows:
\begin{equation}
\label{eq:pslf3a}
\mathbf{A}' = \mathbf{A} +\mathbf{N}_\text{m} - \mathbf{N}_\text{c}.
\end{equation}
where $\mathbf{N}_\text{m} = \mathbf{I}((i,j) \in \mathcal{C}_\text{mustlink})$,
$\mathbf{N}_\text{c} = \mathbf{I}((i,j) \in \mathcal{C}_\text{cannotlink})$, with $i$ and $j$ as indices of a pair of shapes,
$\mathcal{C}_\text{mustlink}$ and $\mathcal{C}_\text{cannotlink}$ collect the sets of shape pairs
with must-link and cannot-link constraints, respectively, and $\mathbf{I}$ is an indicator matrix.

To conduct constrained clustering again using the PSLF framework, we perform another non-negative factorization over the modified similarity matrix:
$\min ||\mathbf{A}' - \mathbf{YS}\mathbf{Y}^T||_F^2$,
where $\mathbf{S} \in{\mathbb{R}^{C \times C}}$ contains cluster centers and
${\mathbf{Y}} \in{\mathbb{R}^{N \times C}}$ is cluster indicator.


\paragraph{Unsupervised style clustering}
\label{subsec:uncluster}

Having computed the fused feature matrix $\mathbf{V}$ in Section~\ref{subsec:fusion},
performing unsupervised style clustering is straightforward.
To do so, we utilize the self-tuning spectral clustering~\cite{zelnik2004self}.
This method produces the state-of-the-art clustering results while determines the number of clusters automatically.
However, directly clustering the fused features may not generate the optimal results since
the per-view features, computed with random patches, may not be the most relevant.
To this end, we devise an iterative algorithm that interleaves clustering and cluster-guided patch re-selection,
which will be discussed in the next subsection.
In fact, such iterative cluster improvement can be also performed in semi-supervised analysis,
through imposing the user constraints in every clustering.

\subsection{Cluster-guided style patch selecting}

The PSLF clustering has been so far based on the patches pre-selected by plain clustering without feature selection (Section~\ref{subsec:patch}).
In fact, PSLF clustering couples feature selection and more importantly, incorporates the user constraints in the semi-supervised setting,
making it both objectively informative and subjectively desirable.
Therefore, it is preferable to use the PSLF clustering to guide a re-selection of mid-level patches,
leading to more discriminative patches, specifically tuned for the unsupervised or semi-supervised tasks.
The re-selected patches can in turn be used to update the PSLF clustering, via further purifying the clusters.

Based on the PSLF clustering results, we re-select discriminant mid-level patches for each view, to be those which are frequent only within one cluster~\cite{xu2014organizing}.
For each style cluster $\mathcal{C}_l$,
we define the support weight of shape $i$ as $\left( {{\omega_{li}}} \right)_{i = 1}^n$, that measures the support of shape $i$
to any mid-level patch.
A mid-level patch is determined as frequent if its weighted sum of support, denoted by discriminant score $\delta_{lj}$,
is greater than a threshold $\delta_l^t$:
\begin{equation}
\label{eq:reselect}
{\mathcal{K}_l} = \left\{ {{j}{\rm{|}}{{\rm{\delta }}_{lj}} > {\rm{\delta }}_l^t} \right\},\;{\rm{where\;\;}}{{\rm{\delta }}_{lj}} = \left| { \sum_{i = 1}^n {\omega _{li}}\left( {2{x_{ij}} - 1} \right)} \right|,
\end{equation}
and $x_{ij}$ is an indicator function showing that shape $i$ supports patch $j$.
If shape $i$ belongs to $\mathcal{C}_l$, weights $\omega_{li}$ are positive, otherwise, they are negative. 
The discriminant score favors a patch that is frequent in cluster $\mathcal{C}_l$ and penalizes its occurrence in other clusters.
Therefore, the patches in $K_l$ are frequent mainly within cluster $\mathcal{C}_l$ that is regarded as discriminant.
Specifically, we define ${\omega _{li}} = {x_{li}}/{C} - 1/N_p$, where
${x_{li}} = I(i \in \mathcal{C}_l)$ with $I(\cdot)$ being a 0-1 indicator function and $\delta_l^t = \mu N_p / C$ holds.
$C$ is the number of clusters where $N_p$ is the total number of patches.
We use $\mu = 0.07$  for all the datasets we have tested.
The final set of patches takes the union of pre-cluster discriminant patches:
$\mathcal{K} = \bigcup_{l=1}^C{\mathcal{K}_l}$.

After the mid-level patch re-selection, we repeat the process of per-view feature extraction, feature fusion for all 3D shapes
and unsupervised or semi-supervised PSLF clustering.
This cluster-and-select process iterates until the clusters and patches become stable.
The final result comprises of purified style clusters, together with a set of style-characterizing mid-level patches,
or \emph{style patches}.

\begin{figure}[tb]
 \centering
  \includegraphics[width=.99\linewidth]{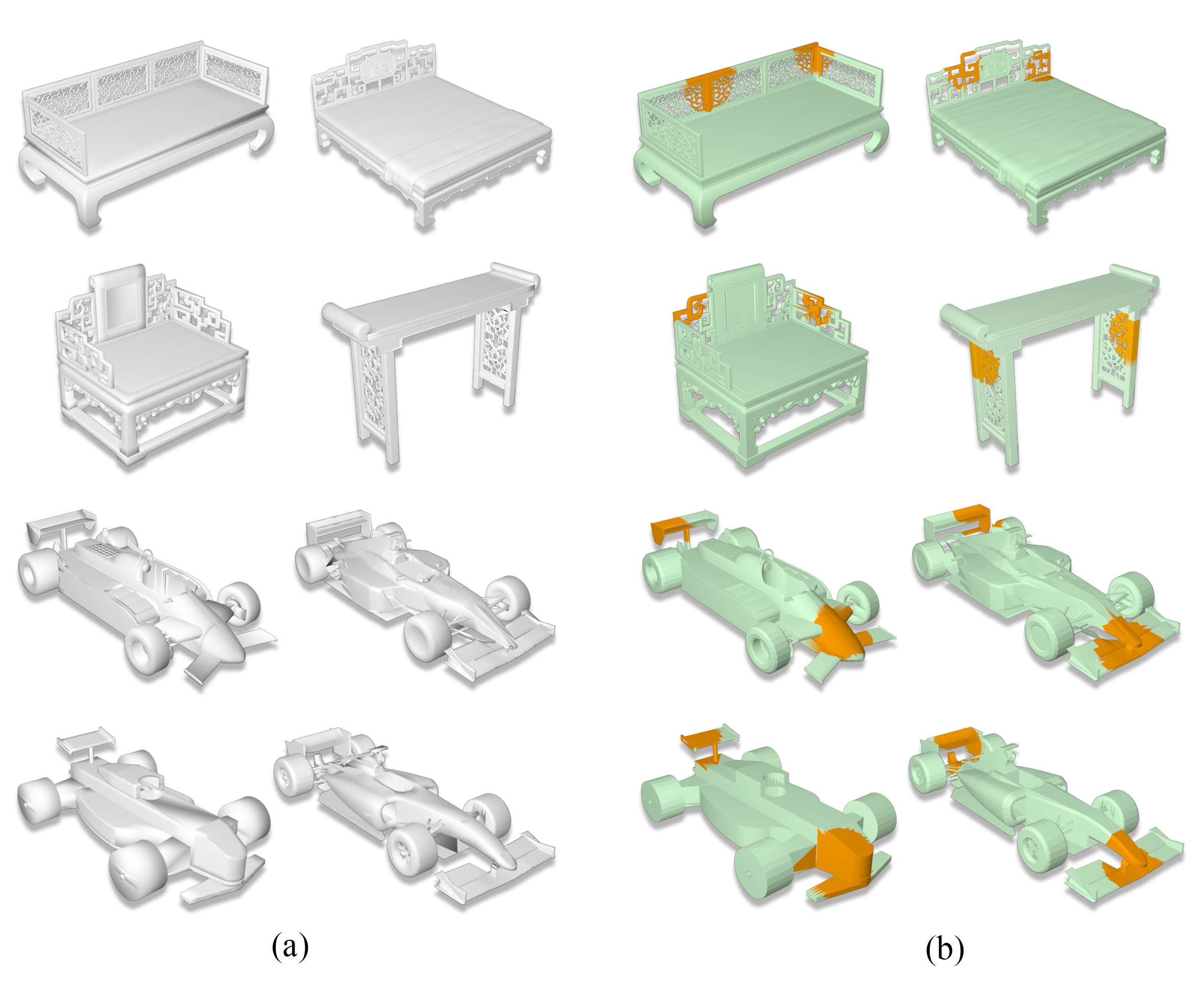}

  \caption{\label{fig:StylePatches}
          Style patches on shape surfaces.
          (a) The input shapes.
          (b) The orange areas are the style patches.}
\end{figure}

\paragraph{Style patch extraction on shape surfaces.}
\label{subsec:style}
One of our goals is to extract style patches on 3D shape surfaces based on the co-analyzed style patches in 2D.
This can be done by backprojecting the 2D patches onto 3D surfaces.
With the 3D patch sampling and projection scheme in Section~\ref{subsec:patch},
we can easily locate the surface region corresponding to a 2D patch.
Note, however, the final style patches are selected from only a few 3D shapes, but not all.
To locate the style patches on other shapes, we need to compare them against the patches sampled from those shapes,
based on HOG features.
Finally, we sort sampled areas on the 3D shape to find style patches according to the number of style patches back-projected to them.
Figure~\ref{fig:StylePatches} shows a few input shapes with the style patches highlighted in orange color.
We also conducted a user study to verify the validity of our detected style patches in Section~\ref{sec:results}.

\begin{figure*}[tb]
  \centering
  \includegraphics[width=.99\linewidth]{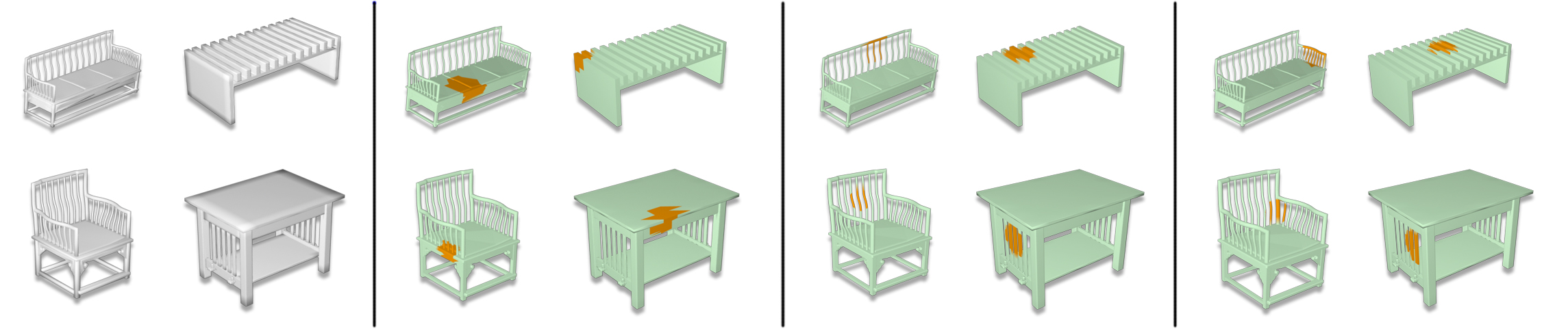}
  \caption{\label{fig:userstudysample}
  A sample query for our user study for style patch localization. (a) Input shapes.
(b) Randomly selected patches. (c) Expert-annotated style patches; (d) Patches returned by
our style analysis method.}
\end{figure*}

\section{Results, evaluation, and applications}
\label{sec:results}

In this section, we first introduce our dataset and then demonstrate experimental results.
Our analysis method is evaluated extensively over a collection of 3D models in
six categories, as shown in Table~\ref{tb:dataset}. A set of comparisons are devised to examine the
effect of our algorithmic components. Our method is also compared
with the state-of-the-art approaches to style analysis. Finally, we demonstrate two
applications that exploit the spatial localization feature of our method.
Extended results of evaluation and comparison, and the full user study data can be found in
the supplementary material.

\textbf{Datasets.}
Datasets employed in our experiments have mainly been collected from the Internet \rz{(e.g. ShapeNet and Trimble 3D Warehouse)}
and previous published works. These datasets include a total of around 2,600 three-dimensional
models arranged into six collections: Mixed Furniture 1, Mixed Furniture 2, Building, Chair,
Car, and Vase. Each collection contains models with multiple styles, where for models sharing the
same style, their geometry and structure can be quite different. In addition, even when all models fall
under the same general category (e.g., chairs), their styles, geometries, and structures can be significantly different.
Note that our method only assumes that the 3D objects have been upright oriented,
but not necessarily consistently aligned.

\textbf{Style labels.}
Our selection of style labels are based on common or professional knowledge
accessible from publications and human experts. For example, furniture styles include
``Simple Chinese'', ``Noble Chinese'', ``European'', ``Country'', and ``Modern'',
according to their decorative styles~\cite{Morley1999The}. Buildings are labeled based
on their geographic-temporal styles such as ``Gothic'', ``Greek'', ``Byzantine'', and ``Asian''~\cite{lun2015elements}.
In the final step, all the style labels for all object collections have been validated by four experts, who
are professors from industrial engineering and architectural design.
Table \ref{tb:dataset} shows the number of styles.

\begin{table}[tb]
\centering
\begin{tabular}{l|r|r}
Shape collection         & \#Shapes     & \#Style classes   \\ \hline\hline
Mixed Furniture 1         & 120                       & 4         \\ \hline
Mixed Furniture 2         & 400                       & 5         \\ \hline
Building                  & 329                       & 4         \\ \hline
Chair                     & 516                       & 9         \\ \hline
Car                       & 1,050                       & 6         \\ \hline
Vase                      & 194                       & 5         \\ \hline
\end{tabular}
\caption{\label{tb:dataset}3D object collections for our style analysis.}
\end{table}

\textbf{Ground truth for style clustering.}
We asked three of the experts to assign each 3D model in our dataset to one of the
available style classes, based on the style labels obtained as described above, for all six
object collections. After that, we asked the fourth expert to verify the style
assignments. A few iterations were performed to arrive at the final labeling,
which serves as the ground truth for style clustering of the models, per object collection.
All the models and style labels can be found in the supplementary material.
\rz{Note that these style clusterings are constructed only for evaluation; they are not
used as training data for our method.}

\textbf{Parameters.}
There are five tunable parameters in the PSLF optimization. Specifically, the smoothness
of different view weights is controlled by $\lambda$; non-negative parameter $\beta$ is to
trade off between the objective of non-negative reconstruction and the $l_{2,1}$-norm regular
item; the weight of the $l_{2,1}$-norm regular item is tuned by $\gamma$; common latent factor
space shared between multiple views is controlled by  $\eta$, where $0<\eta<1$, and finally,
$\varPi=(\pi^1,\pi^2,\ldots,\pi^p)$ controls the weights of different views for all models.
All the experiments have been conducted with a fixed parameter setting:
$\beta\approx 0.05$,
$\lambda\approx 20$, and
$\gamma\approx 10$, and $\eta=0.2$ in PSLF; 
view weights are set as $\varPi= (\frac{1}{P}, \frac{1}{P}, \ldots, \frac{1}{P})$ with $P$ views.
\rz{We also examined the effect of different patch sizes on the purity of style clustering and
found that a size of $48\times48$ generally leads to the best results overall; varying the
patch sizes did not affect the purity more than 5\%. Thus, we fixed the patch size to $48\times48$
in all of our experiments.}
%


\subsection{Style analysis results and evaluation}
\label{subsec:eval}

We first evaluate the performance of our method for style clustering and style patch localization.
Since it is difficult to collect consistent ground truth data for style patches, we instead conduct a
user study where human participants are asked to judge the results produced by our algorithm.
For style clustering with style labels set up, we evaluate our results using the standard
clustering {\em purity\/} measure.
Let $C$ be the set of clusters from a clustering result for a dataset, and let
$L$ be the set of ground truth clusters. For any cluster $c \in C$, its precision against
a ground-truth cluster $l \in L$ is defined as,
$P\left( {c,l} \right) = \frac{{\left| {c \cap l} \right|}}{{\left| c \right|}}$. The purity measure
reflects an average of weighted precision for each cluster and is defined as:
\begin{equation}
\label{eq:purity}
\rho\left( {C,L} \right) =  \sum_{c \in C} \frac{{\left| C \right|}}{{\left| N \right|}}max\left(P\left( {c,l} \right)\right)
\end{equation}
Note that purity depends on its relative maximum precision on ground truth and therefore
it can comprehensively reflect classification or clustering precision.

\textbf{Style patch localization.}
To verify that the final patches returned by our style analysis indeed represent shape styles, we
conduct a user study which compares our results to those annotated by human experts. We
randomly selected $20$ small sets of models from the six object collections, where each set
consists of models belonging to the same style class based on our ground truth. For each
set, we asked human experts to identify style-defining patches by painting over the shapes.
We then conducted a user study where participants are provided with three types of
patches (in color) for the same model set: randomly selected patches, expert-annotated patches,
and patches returned by our style analysis algorithm. Figure~\ref{fig:userstudysample} shows
a sample query for one of the $20$ model sets. Note that in the study, the three choices were
randomly ordered in each query. Each subject is asked to choose which of the three choices
would best reflect the style of the set of models shown. In the study, $20$ model sets were
presented to $58$ human participants with different backgrounds and no prior knowledge
about our method.

\begin{figure}[tb]
\centering
\includegraphics[width=.99\linewidth]{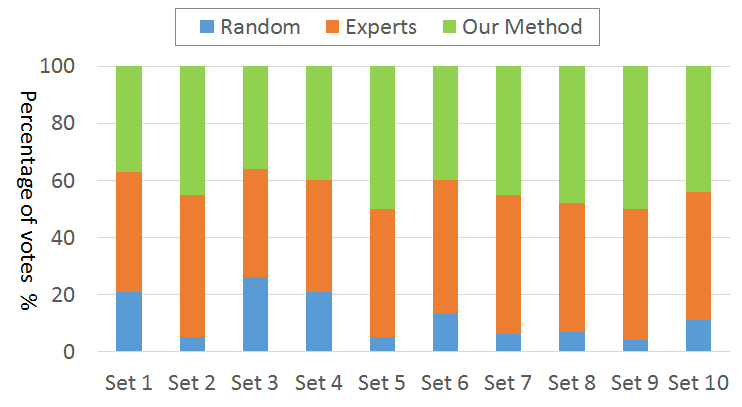}
\caption{Percentage of user votes for three types of patches: random (blue),
expert-annotated (orange), and those returned by our method (green).}
\label{fig:UserStudyStylePatches}
\end{figure}

A subset ($10$ out of $20$ model sets) of results from this study are plotted in Figure~\ref{fig:UserStudyStylePatches}, where the percentage of the user selection of each choice
is shown. The full set of results can be found in the supplementary material.
As indicated in Figure~\ref{fig:UserStudyStylePatches}, participants' preference of our method
is fairly close to that of the expert annotations.

%
%

\begin{figure}[tb]
  \centering
  \includegraphics[width=.90\linewidth]{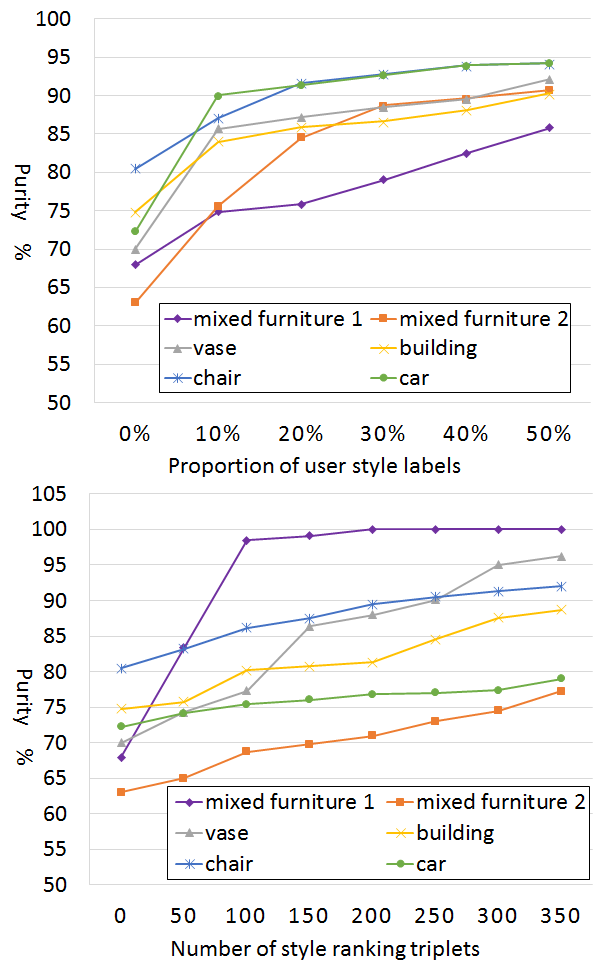}
  \caption{\label{fig:changeofpurities}
  Purity of unsupervised and semi-supervised PSLF clustering as user inputs increase. Top: user input
  as style labels, with unsupervised clustering corresponding to 0\%. Bottom: user input
  as style ranking triplets, with unsupervised clustering corresponding to a count of 0.}
\end{figure}

\textbf{Semi-supervised analysis.}
Our semi-supervised style analysis accepts two types of user inputs: style labels or style
ranking triplets. Figure~\ref{fig:changeofpurities} shows how the style clustering performance,
in terms of purity (see Equation \ref{eq:purity}), changes as we increase the percentage of user labels or
the number of user-specified ranking triplets, respectively. All results were obtained with
iterative PSLF-based clustering. As expected, clustering improves as user inputs increase.
However, the improvement appears to level off when the label percentage passes 30\% or
the number of triplets reaches around 300. Note that 300 only represents an extremely
small number of triplets out of the total number of triplets for an object collection.


%
%

%
%

\textbf{Parameter analysis.}
\rz{We examine two key parameters, view count and $\eta$, which have the most significant impact
on results among all the parameters and they need to be carefully selected.

%
\begin{figure}[htb]
  \centering
  \includegraphics[width=.99\linewidth]{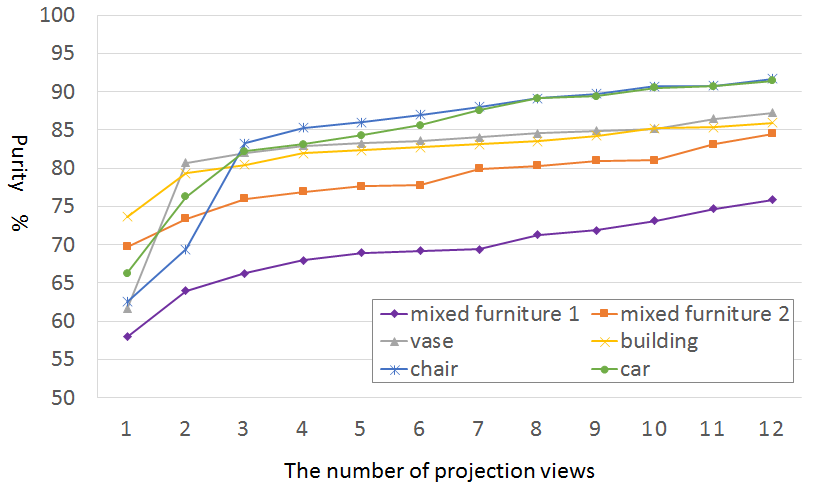}
  \caption{\label{fig:viewsimpact}
          Selection of the impact of shape projection views on our method.}
\end{figure}

We tested our style clustering with different number of views, ranging from $1$ to $12$ views.
For a given number of view, we enumerate all different combinations of views, picked from the full set of $12$ views.
The final results for each view count were averaged over all different combinations.
As shown in Figure~\ref{fig:viewsimpact}, the clustering results gradually improves with increasing views.
The typical "leveling out" points appear to be 10-12 views. 

We have also analyzed the impact of parameter $\eta$ on classification accuracy for all datasets, in Figure~\ref{fig:Parameter_eta}.
The parameter controls the proportion of common latent factor space shared across different views in PSLF.
It can be observed that $0.2$ gives the best results.
This also verifies that PSLF is well-suited for our problem when both the consistency and complementarity
of different views are exploited. Thus, we fix $\eta=0.2$ in all experiments throughout the paper.}

\begin{figure}[tb]
  \centering
  \includegraphics[width=.99\linewidth]{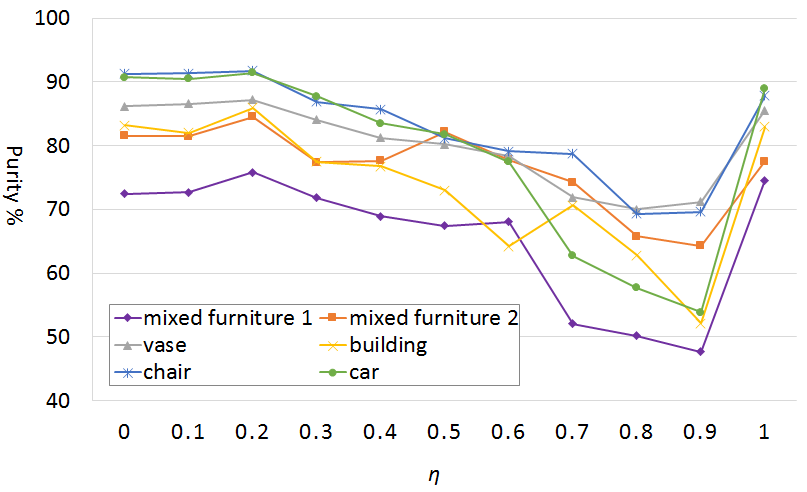}
  \caption{\label{fig:Parameter_eta}
  Classification accuracy as functions of $\eta$. We conduct experiments for $\eta$ with 20\%
  labels to illustrate its role in PSLF.}
\end{figure}

\subsection{Comparisons}
\label{subsec:comp}

We now provide several comparative studies to validate important design choices made in
our method and to show how well our semi-supervision performs on the task of style
similarity learning as compared to the state-of-the-art methods.

\textbf{Style clustering: PSLF learning vs.~PCA and CCA.}
We compare our PSLF-based style clustering method with PCA~\cite{jolliffe2002principal}
and Correlation Analysis-based approaches (CCA)~\cite{chaudhuri2009multi}. Both PCA and CCA are dimensionality reduction techniques, just like PSLF.

\begin{figure}[tb]
  \centering
  \includegraphics[width=.99\linewidth]{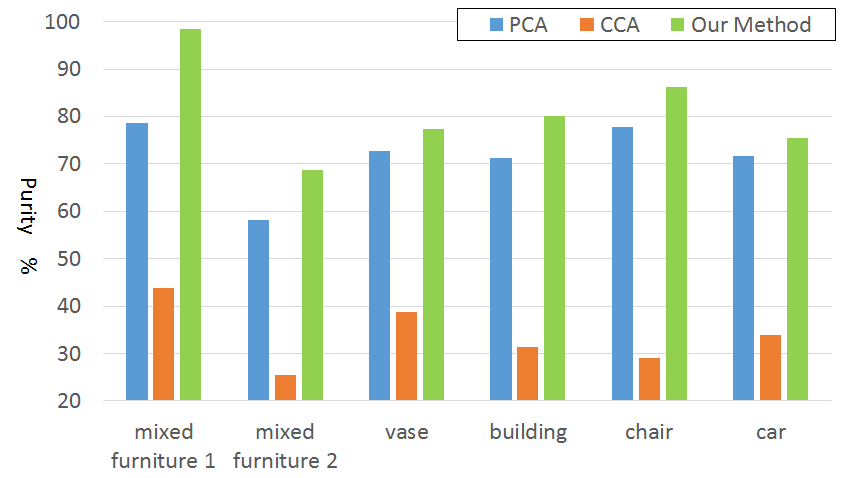}
  \caption{\label{fig:PSLFComparison}
  Comparison on style clustering between our PSLF (green) and other feature fusing methods,
  PCA (blue) and CCA (orange). User input consists of 100 ranking triplets.}
\end{figure}

In the experiment, the reduced dimensionality of PCA is the same as the dimensionality of the
partially shared latent representation in PSLF. CCA is a two-view method; we have executed the
algorithm with each pair of two-view data and report results obtained using the best pair.
The comparison is conducted using triplet constraints (100 triplets) to drive semi-supervised
style clustering (see Section~\ref{subsec:Tripletcluster}, which is based on fused features of
each method for each data set. Figure~\ref{fig:PSLFComparison} suggests that using PSLF to
fuse the features tends to produce higher style clustering purities than PCA and CCA. We believe
that the underlying reason is that PSLF can improve its fused features with the aid of triplet
constraints while PCA and CCA cannot. In turn, the improvement on feature fusion is responsible
for higher-accuracy clustering results. Additional results obtained by changing the amount of
user inputs can be found in the supplementary material.

\begin{figure}[tb]
  \centering
  \includegraphics[width=0.99\linewidth]{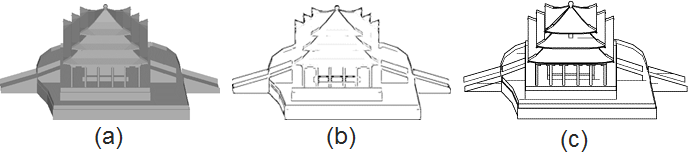}
  \caption{\label{fig:CLD_comparison}
  Different projective methods. (a) Rendered images as ~\cite{su2015multi}. (b) Coherence-enhancing line drawing. (c) Our line drawing.}
\end{figure}

\textbf{Input representation: Line drawings vs.~rendered images.}
To examine the effectiveness of using line drawings as input representation for style analysis,
we make a comparison to the use of rendered images. For the latter, we render a 3D
shape with the same setting as~\cite{su2015multi}, obtaining a set of RGB images
from the same views as those for line drawings.
Besides, we also compare to coherence-enhancing line drawing(CLD)~\cite{Wang2013Coherence} extracted from
the rendered images. Note that CLD, computed based on 2D images, is inherently
different from our projective line drawings which may contain
geometric features of 3D surfaces.
Figure~\ref{fig:CLD_comparison} shows an example shape in the above three different projections.

\begin{figure}[tb]
  \centering
  \includegraphics[width=0.9\linewidth]{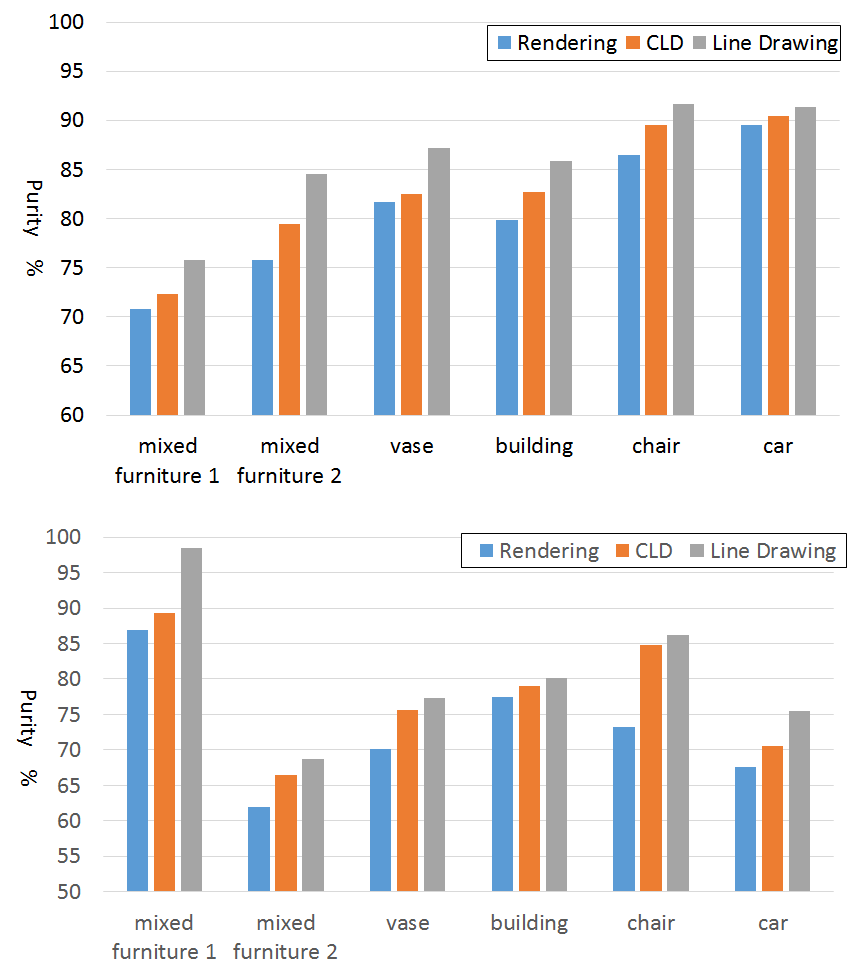}
  \caption{\label{fig:Projective image comparison}
  Semi-supervised style clustering over projected line drawing (orange) vs. rendered images (blue).
  Top: results with 20\% user label constraints. Bottom: results with 100
  style ranking triplets.}
\end{figure}

We conduct experiments on the three types of projection images, under the constraints
of $20\%$ user-prescribed style labels or $100$ style ranking triplets, meanwhile keeping all design
choices (e.g., use of HOG features) and parameters the same. Results shown
in Figure~\ref{fig:Projective image comparison} indicate the consistent superiority of using
projective line drawings for our task.
The advantage over rendered images clearly verifies that feature lines are more directly related to
shape styles. The superiority of our projective 3D line drawings over 2D edge
detections further reflects the importance of 3D feature lines.
Additional results with different percentages of user-prescribed labels
are give in the supplemental material.

\begin{figure}[tb]
  \centering
  \includegraphics[width=0.9\linewidth]{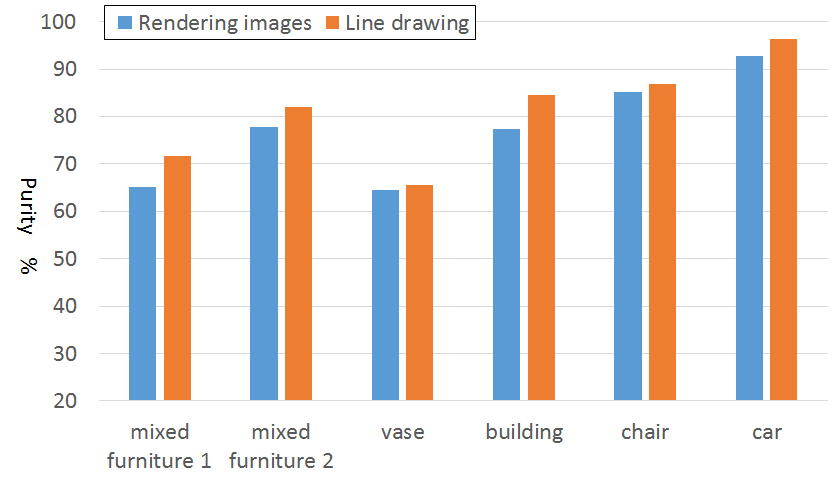}
  \caption{\label{fig:Projective_image_comparison_in_Lenet}
  Comparison on style clustering over projected line drawing (orange) vs. rendered images (blue) in Lenet~\cite{Lecun1998Gradient}.}
\end{figure}

It is known that the neuron activations in the lower level of convolutional neural networks
are able to capture characteristic lines and corners in an image.
Therefore, it is interesting to see how our projective line drawings
compares to rendered images when using CNN for feature learning.
For simplicity, we train a LeNet~\cite{Lecun1998Gradient} ($5$ convolutional layers plus $2$ fully connected layers) with $20\%$ of our datasets, to learn feature representation for each view separately.
The view-wise features are fused with PSLF for final style clustering.
For projective line drawings, the raw input is represented in HOG space.
Rendered images, however, are input directly.
Figure~\ref{fig:Projective_image_comparison_in_Lenet} shows that
line drawing outperforms color rendering, when both using CNN for feature extraction
and PSLF for feature fusion.
This is because projective lines include 3D geometric features which cannot be recovered
by 2D convolutional operations. Moreover, line drawings avoid the effect of lighting condition
which is an inevitable problem for rendering.


\begin{figure}[tb]
  \centering
  \includegraphics[width=0.9\linewidth]{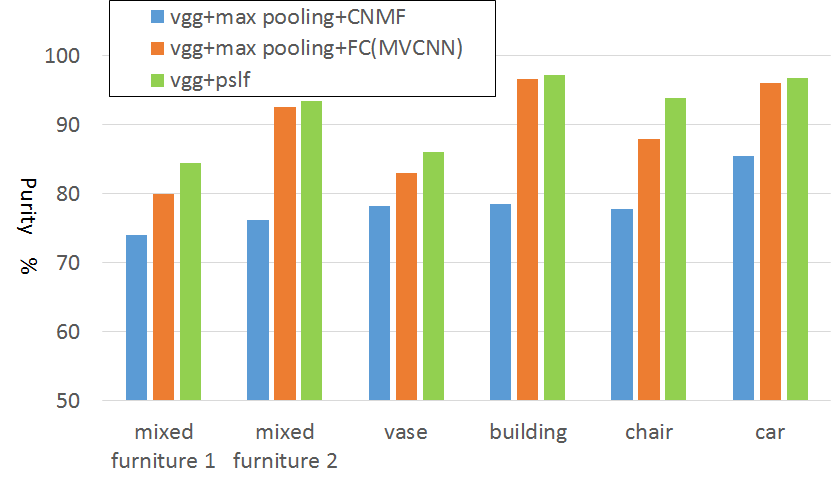}
  \caption{\label{fig:PSLFComparison_in_vgg}
   Comparison on style clustering between PSLF(green), max-pooling with CNMF(blue) and MVCNN(orange). They are performed on the same line drawing features extracted from VGGNet-16~\cite{Simonyan2014Very}.}
\end{figure}

\textbf{Feature fusion: Max pooling vs. PSLF.}
%
A key component of multi-view style analysis is how to fuse the features extracted for the multiple view channels.
A commonly used fusion scheme is to perform max pooling operation over the multi-channel features~\cite{su2015multi}.
However, such a simplistic fusion is oblivious to the consistency and complementarity among the multiple channels
which are both essential to effective feature fusion. To verify this,
we train a VGGNet-16~\cite{Simonyan2014Very} to extract view-wise features from projective line drawings,
and then utilize max pooling and PSLF to fuse the features, respectively.
The dimensionality of the fused feature is $512$ for max pooling and $50$ for PSLF.
For max-pooling features, we employ CNMF~\cite{Liu2010Non} to perform style clustering,
under the constraint of $20\%$ data with known labels, same as PSLF.
We also compare the results to MVCNN~\cite{su2015multi}, using the same $20\%$ data as the training set.
The results in Figure~\ref{fig:PSLFComparison_in_vgg} show that PSLF outperforms
max pooling, both with CNMF and in MVCNN, on all datasets.
This verifies that PSLF, as a clustering method with a carefully designed feature fusion scheme,
is especially suited for multi-view style analysis.

\textbf{Learning style similarity.}
State-of-the-art methods for learning style similarities from style ranking triplets include
the recent works of Lun et al.~\shortcite{lun2015elements}, Liu et al.~\shortcite{liu2015style}
and Lim et al.~\shortcite{Lim2016Identifying}.
Since our semi-supervised PSLF learning also accommodates style ranking triplets as user input
(see Section~\ref{subsec:Tripletcluster}), these methods and our method can be compared
for the task of {\em predicting style similarity rankings\/} using the learned similarity distance.
Our comparisons were conducted on datasets from the two previous works, respectively. As well,
the set of style ranking triplets were also reused from their works. We split the set of
triplets into a training set for learning and a testing set. Prediction accuracy is measured on how
accurate the learned similarity distance would predict the similarity relations among the three
data entities in a testing triplet.

\begin{figure}[tb]
  \centering
  \includegraphics[width=0.9\linewidth]{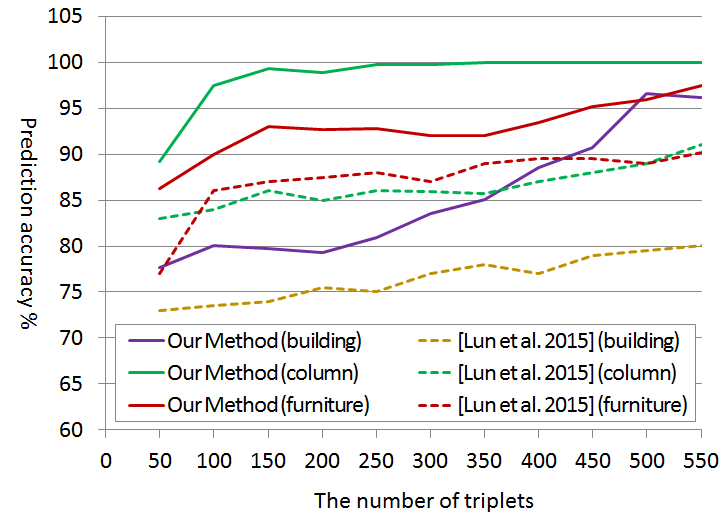}
  \caption{\label{fig:Predictioncomparison}
  A comparison on style ranking prediction accuracy with~\protect\cite{lun2015elements}, over four
  of their object categories.}
\end{figure}

Figure~\ref{fig:Predictioncomparison} shows a comparison to Lun et al.~\cite{lun2015elements}
on four of their seven object categories, where the number of triplets used for training varies from $50$ to
$550$. The remaining three categories had much fewer available triplets and the corresponding
comparison results can be found in the supplementary material.
Our method leads to higher accuracies in all cases with only two exceptions: the lamp and dish datasets. Our performance on the dish set is below that of Lun et al.~\cite{lun2015elements} and
we believe this is due to the fact that the dish shapes are mostly smooth and they lack line-type features,
while our method relies only on features from projected line drawings. On the other hand,
Lun et al.~\cite{lun2015elements} employs a large set of features, including projective ones.
For the lamp set, since it contains a large number of models and variations, it is conceivable that by
relying on a more limited feature set, our semi-supervised learning would require more training to
reach a performance plateau.

\begin{table}[!t]
\centering
\begin{tabular}{l|c|c|r}
\hline
Scene category  & {\cite{liu2015style}}   & {Ours}   & {\%triplets} \\ \hline\hline
living room                   & 73\%          & {\bf 85\%}  & 10\% \\ \hline
dining room                 & 72\%          & {\bf 74\%}  & 30\% \\ \hline
\end{tabular}
\caption{\label{tab:comparison}Comparing style ranking prediction accuracy with~\protect\cite{liu2015style} on their
scene datasets. The last column shows $\%$ of style ranking triplets employed by our method
for training as opposed to \protect\cite{liu2015style}.}
\end{table}

\begin{table}[!t]
\centering
\begin{tabular}{l|c|c}
\hline
Category     & {~\cite{Lim2016Identifying}} & {Ours} \\ \hline\hline
building            & 88.8\%              & \textbf{96.1\%}          \\ \hline
coffee set          & 89.2\%               & \textbf{93.24\%}          \\ \hline
column               & 98\%              & \textbf{100\%}          \\ \hline
cutlery             & 81.2\%              & \textbf{96.39\%}          \\ \hline
dish                & \textbf{90.8 \%}             & 82 \%         \\ \hline
furniture            & 86.2 \%             & \textbf{97.47\%}          \\ \hline
lamp                & 88.5 \%             & \textbf{100  \%}        \\ \hline
\end{tabular}
\caption{\label{tb:dlcomparison}
A comparison on style ranking prediction with~\cite{Lim2016Identifying}, over seven datasets from their paper.}
\end{table}

We also compare style ranking prediction accuracy with~\cite{liu2015style} using the two
scene datasets tested in their paper, as shown in Table~\ref{tab:comparison}. We do not use
all the ranking triplets as in~\cite{liu2015style}. Instead, we randomly sample a subset. As
can be observed, with a relatively small percentage of triplets employed for training, our
method is able to achieve comparable or better prediction accuracies.

In~\cite{Lim2016Identifying}, the authors propose to identify the style of 3D shapes
based on deep metric learning.
Their evaluations are performed on the same datasets as Lun et al.~\shortcite{lun2015elements}. Table~\ref{tb:dlcomparison} reports a comparison to~\cite{Lim2016Identifying}
on their seven object categories with the same triplets from~\cite{lun2015elements} for training.
Specifically, $550$ triplets were used for the building, column, furniture and lamp set,
$150$ for the coffee set, and $100$ for the cutlery and dish set.
Their model was trained with only 3D shapes; no photo was used.
Our method outperforms theirs for all sets except the dish one.
The reason is that most dish models do not contain rich line features.

\textbf{Style classification.}
\rz{We compare style classification results obtained using our method with those from the recent supervised
method of Hu et al.~\cite{Style17}, over the five datasets used in their work:
1) Furniture ($618$ models in $4$ styles);
2) Furniture legs, ($84$ models and $3$ styles);
3) Buildings ($89$ models in $5$ styles);
4) Cars ($85$ models in $5$ styles);
5) Drinking vessels ($84$ models in $3$ styles).

Similar to their work, we run a classification experiment with a 10-fold cross-validation.
However, there is a difference.
They learned the sets of style-defining patches to represent shapes
and train $k$NN classifiers for each style on $9$ folds.
We, instead, used $9$ folds as style labels in our method.
We evaluate the classification accuracy on the remaining fold for both methods.
Finally, we compute the average accuracy for the $10$ folds for all style labels in each set,
based on the ground-truth labels provided in Hu et al.~\cite{Style17}; the results are
shown in Figure~\ref{fig:Classification_comparison}. We can observe that our method
outperforms theirs for all object categories except for buildings.

\begin{figure}[tb]
  \centering
  \includegraphics[width=0.9\linewidth]{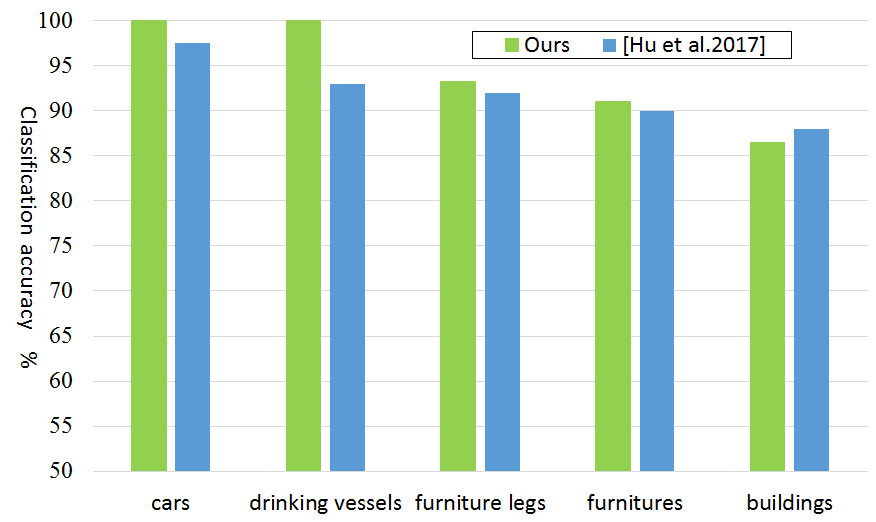}
  \caption{\label{fig:Classification_comparison}
  \rz{Comparison on style classification between our method (green) and Hu et al.~\protect\cite{Style17} over
   datasets from their work.}}
\end{figure}

%
%
%
\rz{We believe that the general improvements over Hu et al.~\cite{Style17} on this task may be
attributed to two factors.
First, projected line drawings and their associated features may be more suitable for the kind of decorative
styles and man-made shapes tested. Shape features considered in Hu et al.~\cite{Style17} could be
severely degraded due to low mesh resolution, tessellation quality, and other geometric imperfections, while
line drawings are most robust against these issues.
Second, their method relies on mid-level 3D patches, which would only capture local geometry information.
In contrast, our method captures both local and global information owing to the convolutional feature encoding
which is known to be more apt at hierarchical feature learning.}}

\textbf{Style patch localization.}
\rz{
Finally, we compare our method with Hu et al.~\cite{Style17} on style patch localization, over their
datasets, the same ones tested above for style classification.
Since their method is supervised with given style clusters, we only compare the feature selection components
of the two methods. That is, our method would take the same style clusters as input to their method.
%

Comparing patch localization is not straightforward since the located patches for each shape is not unique for
either method. To simplify matters, we carry out the comparison on 19 representative style patches obtained
by the method of Hu et al.~\cite{Style17}. These 19 patches come from 19 shapes (one patch per shape)
encompassing all five object categories; they were selected and shown in Figure 11 of the paper.
Hu et al.~\cite{Style17} qualified them as style patches that ``capture {\em distinctive
characteristics\/} of the styles.''
\rz{For each of the 19 shapes, the representative style patch from our method is chosen as the one which is
deemed to be a style patch over the most views.\/}

Figure~\ref{fig:Style_Patches_comparison} shows a visual comparison on 10 out of the 19 shapes. We can observe
that the style patches detected bear some similarities in general, but our style patches tend to be more
feature-rich. The rest of the results can be found in the supplementary material.

\begin{figure}[tb]
  \centering
  \includegraphics[width=0.99\linewidth]{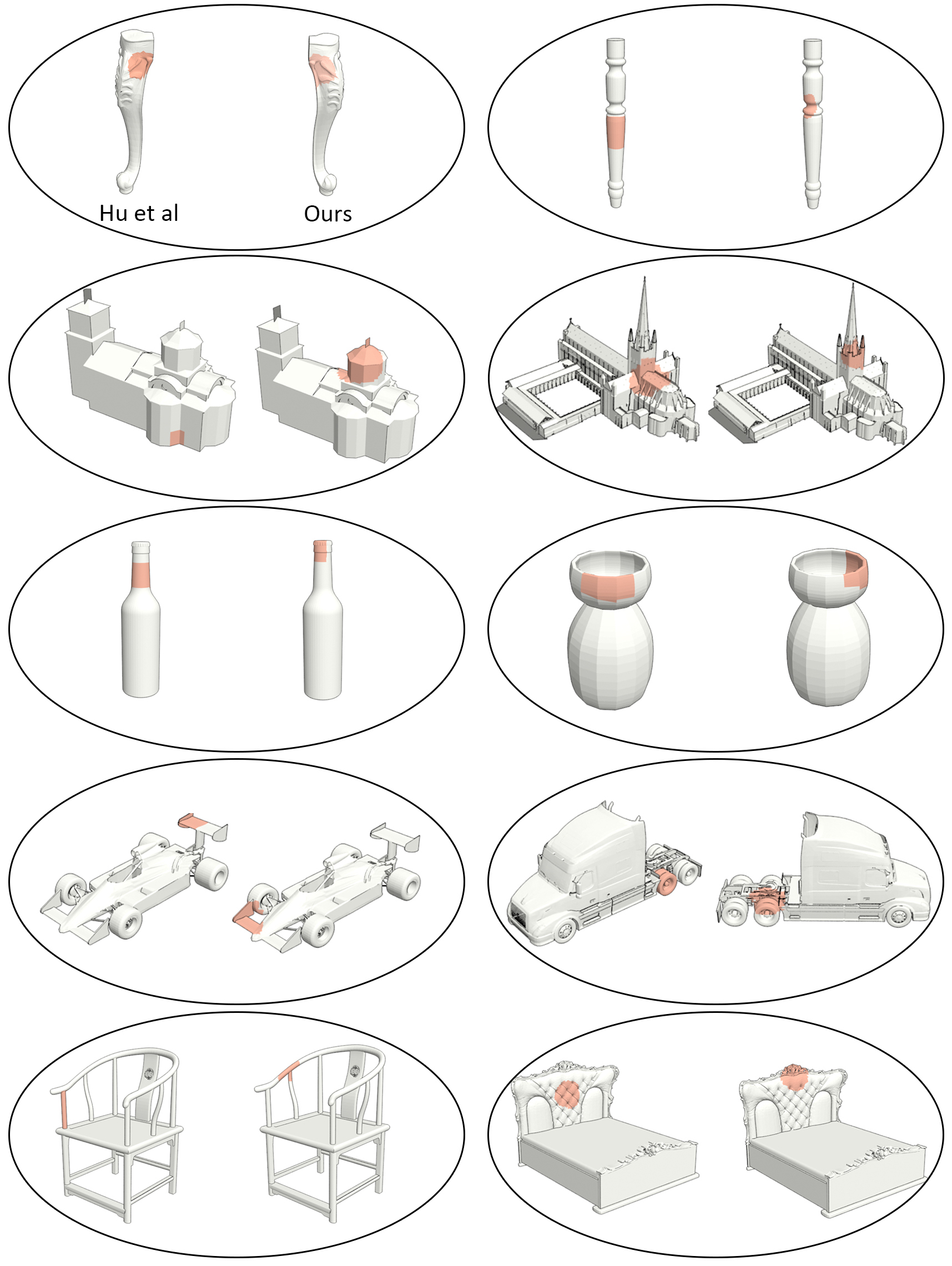}
  \caption{\rz{Visual comparison of style patches located by our method (right one in each pair) vs. those found by \protect\cite{Style17} (left one in each pair).}}
  \label{fig:Style_Patches_comparison}
\end{figure}

\begin{figure}[tb]
  \centering
  \includegraphics[width=0.99\linewidth]{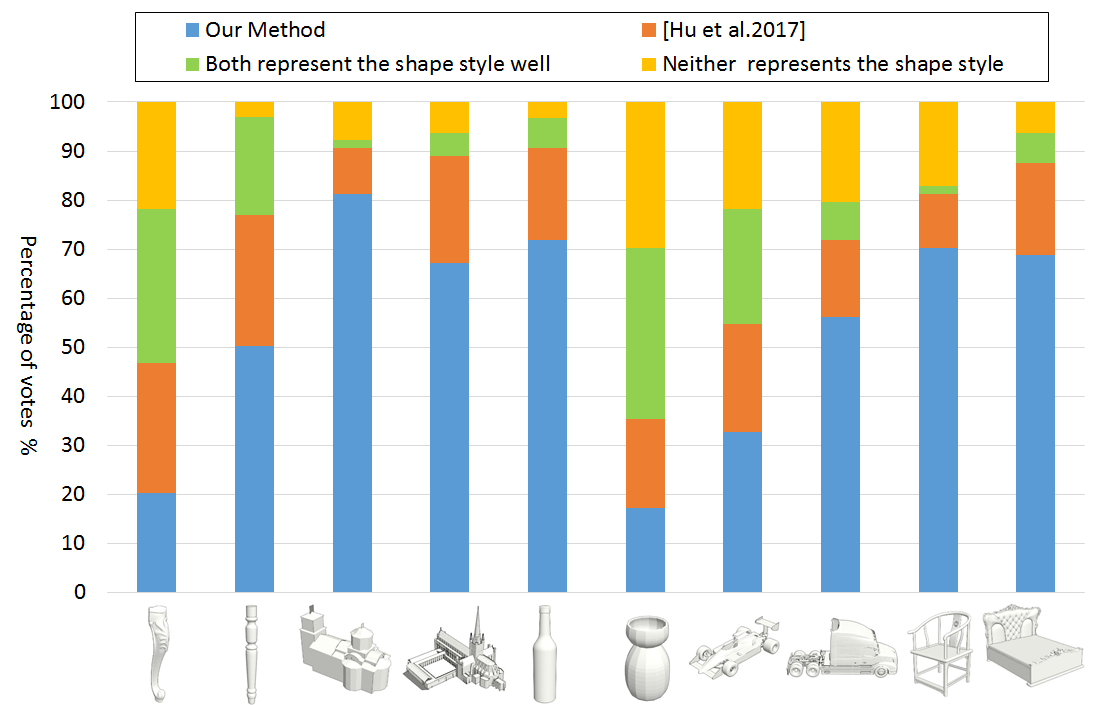}
  \caption{\label{fig:Style_Patches_comparison_user_study}
  \rz{Subset of results from the user study on style patch localization, comparing our method to that of \protect\cite{Style17}. We plot the percentage of user
  choices over the four different answers for 10 shapes.}}
\end{figure}

For a quantitative comparison, we conducted a user study to evaluate the representative style patches found by
the two methods on the 19 shapes. The study consists of 19 queries, one per shape. For each query, subjects were
provided with the two style patches, marked as $A$ and $B$ and shaded in the same color, for the same shape. Then
the subjects were asked to choose one of four possible answers in regards to the style patches:
1) Patch $A$ represents the shape style better;
2) Patch $B$ represents the shape style better;
3) Patches $A$ and $B$ both represent the shape style well;
4) Neither set of patches represents the shape style.
For each test shape, our result and theirs are randomly ordered.

A total of $58$ subjects participated in the study; these subjects are all students from various disciplines including computer
science, architecture, arts, and information management. Results on a subset of ($10$ out of $19$) shapes are shown in
Figure~\ref{fig:Style_Patches_comparison_user_study}, where the percentages of user selections for different answers are
plotted; the remaining results are available from the supplementary material. It is quite evident that style patches extracted by
our method were generally more favored by the human subjects.}

\subsection{Applications}
\label{subsec:app}

\begin{figure}[!t]
  \centering
  \includegraphics[width=.99\linewidth]{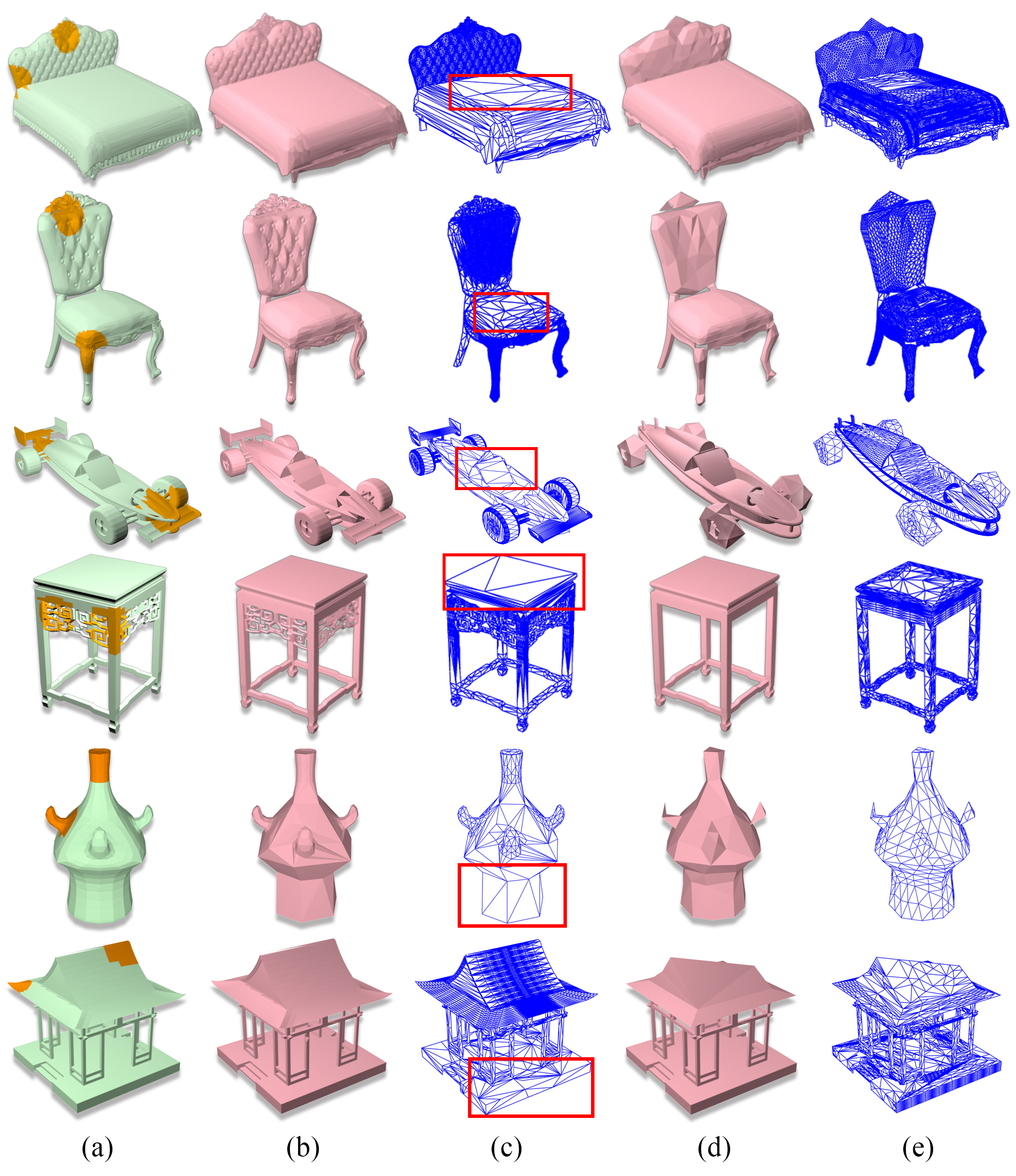}
  \caption{\label{fig:ShapeSimplification}
  Style-aware mesh simplification. (a) Original mesh models with style patches. (b-c) Shaded and wireframe versions of
  simplified models with style preservation via constrained quadric-based decimation; red boxes
  highlight significant triangle reduction near non-style areas. (d-e) Simplified models without style
  preservation, via unconstrained quadric-based decimation.}
\end{figure}

\textbf{Style-aware mesh simplification.}
%
Spatial localization of style patches allows a simple scheme to be developed for
style-aware mesh simplification. The main extra step, after obtaining the style
patches from our current method, is to extend the few style patch {\em samples\/}
returned to entire style regions. Then, we can apply a constrained version of quadric-based
mesh simplification~\cite{Garland1997Surface} while keeping the style regions in tact.
Figure~\ref{fig:ShapeSimplification} shows some results, where the number of triangles
(after 70\% reduction) are the same for the simplified models with and without preserving
styles. Apparently, style-aware simplified models better resemble original models style-wise
and have larger triangles over flat areas highlighted by red boxes.

For the patch-to-region extension, all we need to do is to compare the initially sampled patches
(see Section~\ref{subsec:patch}) with the detected style patches and then mark all those with HOG-space
similarity above a threshold as stylistic. All the style patches are finally back-projected to
the 3D shape to aggregate into style regions over the shape. To improve style region boundaries,
we re-sample three times as many initial patches as before to increase the resolution.

\begin{figure}[!t]
  \centering
  \includegraphics[width=.99\linewidth]{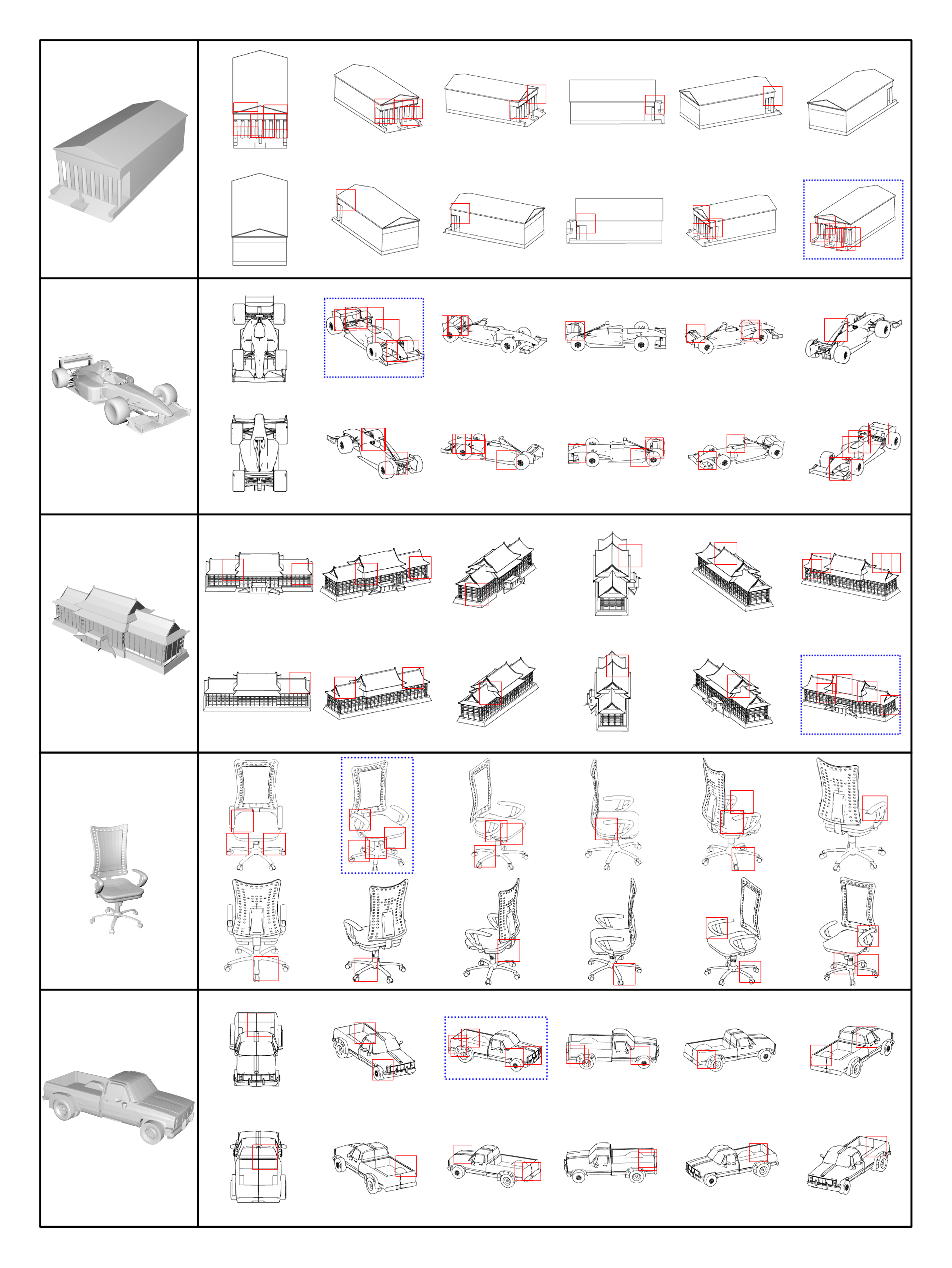}
  \caption{\label{fig:ViewSelection}
  Style-aware view selection. For each 3D model, we show it, as projected line drawings, in the 12 views employed by our multi-view style analysis method.
  Detected style patches and those deemed to be similar to them are shown in red boxes. The best view, one with the most red boxes, is highlighted by a blue
box, and also shown in the left column.}
\end{figure}

\begin{figure}[!t]
  \centering
  \includegraphics[width=.99\linewidth]{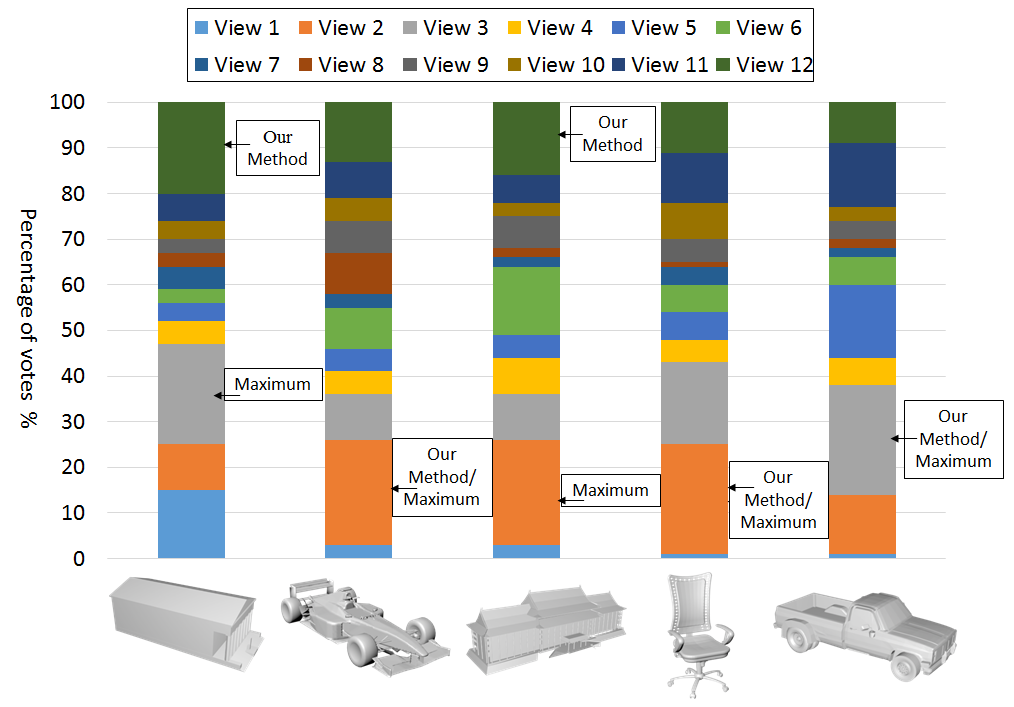}
  \caption{\label{fig:UserStudyViewSelection}
  Results of the user study on best view selection. Percentage of user votes for each view
  is shown. Views selected as best by most human subjects and views selected by our style-aware
  selection scheme are both indicated. At the bottom, we show best views selected by human
  subjects, which can be contrasted with the best views found by our method in Figure~\ref{fig:ViewSelection}.}
\end{figure}

\textbf{Style-aware view selection.}
Another application for spatial localization of style patches is style-aware view selection,
where views that are deemed to be most {\em style-revealing\/} for a 3D object are identified.
After style patches are obtained as discussed in Section~\ref{subsec:style},
we use them to first detect similar (initially sampled) patches in HOG space for each
considered view. We then select the view that has the maximum number of
patches that are sufficiently similar to the style patches. Figure~\ref{fig:ViewSelection}
shows some of the best object views obtained this way, as opposed to other views.

We compare our style-aware best view selection with human judgment.
For a test 3D model, we let 58 human subjects select, among the 12 views employed in our
multi-view style analysis, which one provides the ``best view" for the model. In
Figure~\ref{fig:UserStudyViewSelection}, we show the percentage of subject votes for each view,
for five selected 3D models. The results of this study reveal that our simple best view selection
based on extracted decorative shape styles tends to obtain the same or close views as the
human subjects would.

\begin{figure*}[!t]
  \centering
  \includegraphics[width=.99\linewidth]{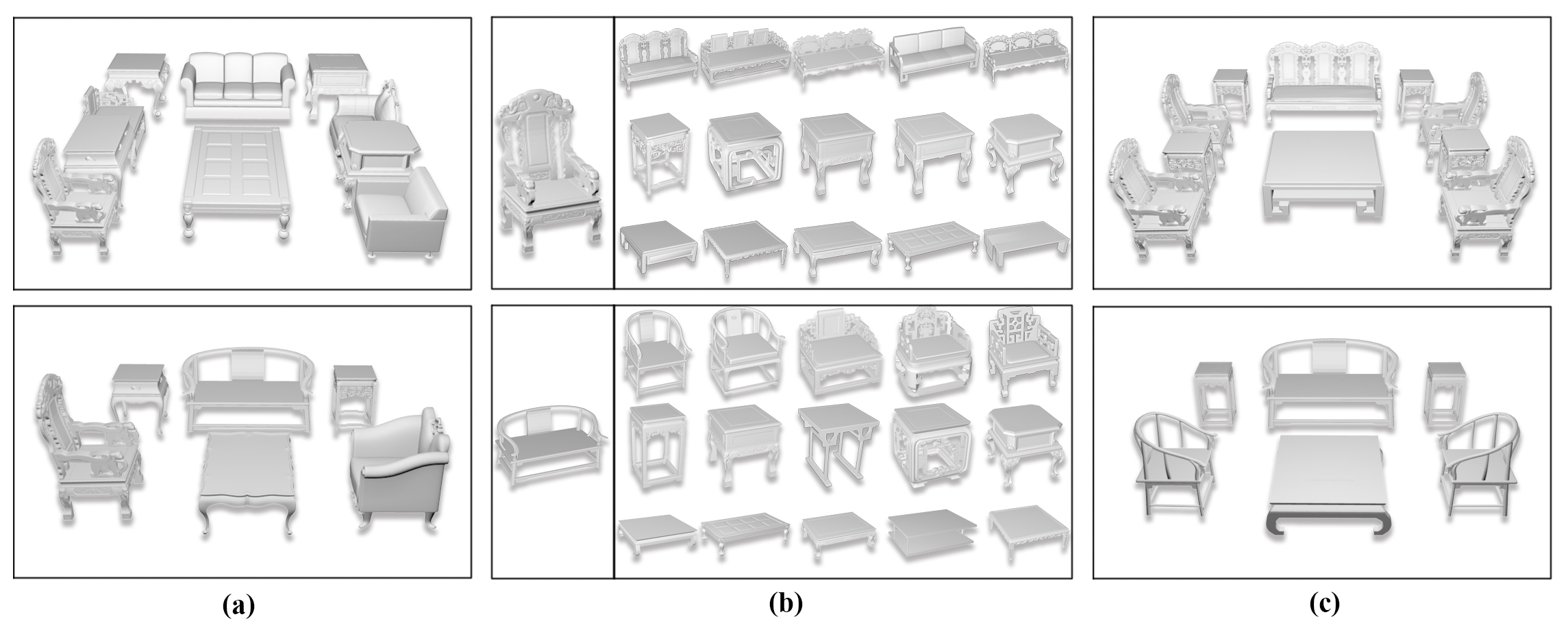}
  \caption{\label{fig:furniture_recomm.png}
  Style-aware furniture recommendation.
  (a) Input scene consisting of objects with varying styles.
  (b) One piece of furniture from the input scene as query to retrieve stylistically similar pieces from different furniture categories.
  (c) The recommended, stylistically closest, furniture pieces replace existing pieces in the scene to achieve style compatibility througout.}
\end{figure*}

\textbf{Style-aware furniture recommendation.}
Style similarities resulting from our analysis can enable a style-aware furniture
recommendation application, much like the one developed in~\cite{liu2015style}. Specifically,
furniture pieces can be retrieved from a shape repository to populate a scene based on
style similarity (see Figure~\ref{fig:furniture_recomm.png}).

\begin{figure}[!t]
  \centering
  \includegraphics[width=\linewidth]{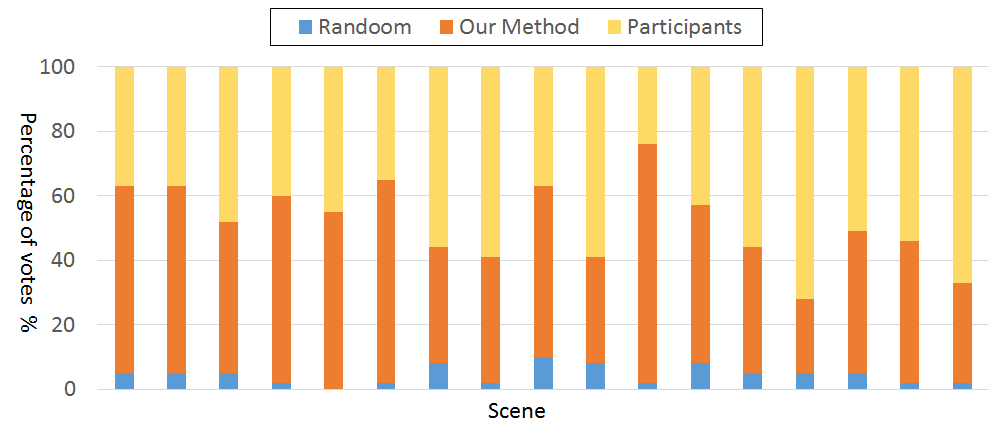}
  \caption{\label{fig:style_sim_study}
  User study results comparing style-aware furniture recommendation via random selection (blue), human judgment (yellow), and
  our style similarity measure (orange). The bars show the percentage of votes from human participants.}
\end{figure}

To verify the effectiveness of the style similarity measure obtained from our work, we conduct a user study where
human participants were asked to choose the scene containing furniture pieces that possess the highest level of style compatibility among a triplet of 3D scenes. In each scene, one of the furniture pieces
is recommended. Among the triplet, one scene is randomly arranged,
one has the recommendation given by human users, and one has the
recommendation selected as the furniture piece whose style is the closest match with other
furniture pieces in the scene. The study consists of 17 scene triplets, all randomly sorted, and
58 human participants to make scene selections.
Figure~\ref{fig:style_sim_study} shows the scene selection results from the user study, demonstrating that
style-aware furniture recommendation using our style similarity performs comparably with
human recommendation. All the scenes used in the study can be found in the supplementary material.

\begin{figure}[!t]
  \centering
  \includegraphics[width=\linewidth]{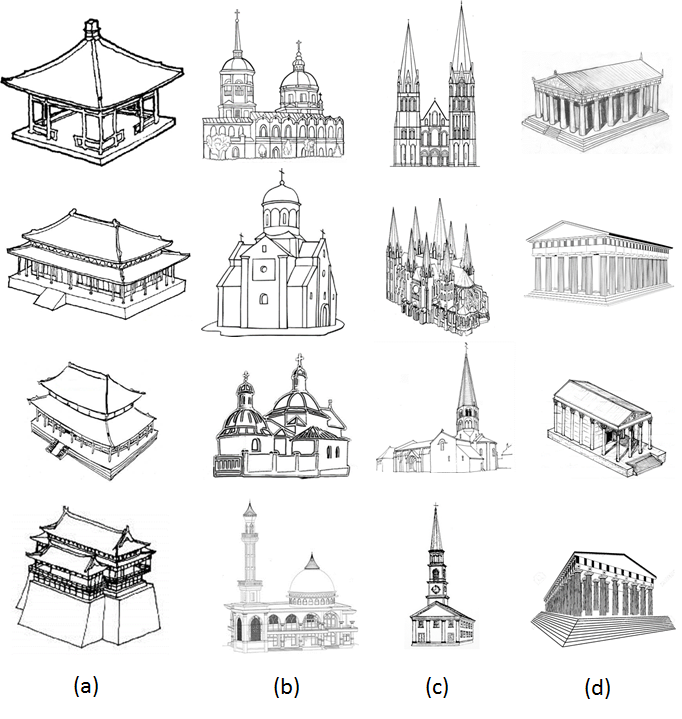}
  \caption{\label{fig:2D_train_data}
  Examples of building sketches from Internet. (a) Asia. (b) Byzantine. (c) Gothic. (d) Greece.}
\end{figure}

\textbf{3D architectural style identification based on 2D sketches.}
Architectural styles are commonly documented with the help of 2D sketches in professional architecture books.
Figure~\ref{fig:2D_train_data} shows $16$ examples of $160$ building drawings collected from the Internet,
in four styles: Asia, Byzantine, Gothic and Greece.
More drawings can be found in the supplemental material.
It is sometimes desirable to identify the architectural style of a 3D building based on
professional 2D drawings.
The projective learning nature of our method enables such a cross-modality style recognition task.
We use our method to select the representative style patches and extract feature for each 2D sketch.
Given a test 3D shape, we project it into $12$-view line drawings and find the closest 2D sketch for each view,
based on Euclidean feature distance.
Each view casts a vote to its closest style and the style receiving the most votes is output.
One can also compare the representative style patches in 2D sketches to the sampled patches on each view based on HOG features, to locate the style patches and back-project them onto the 3D surfaces.

\begin{figure}[!t]
  \centering
  \includegraphics[width=\linewidth]{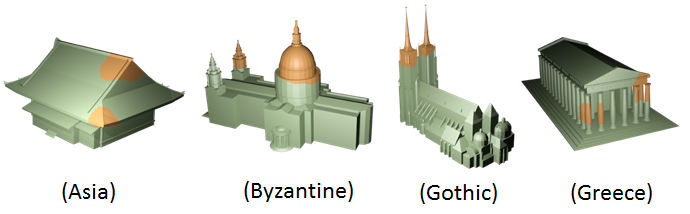}
  \caption{\label{fig:style_localization_app}
  Examples of style prediction(label below shape) and style prediction(orange rendered).}
\end{figure}

We use our building models to test style prediction and style patch location.
A sample of results are shown in Figure~\ref{fig:style_localization_app}.
The style prediction accuracy is $71.4\%$, which is slightly lower than the $74.77\%$ by our unsupervised style clustering with feature fusion. The reasons include: 1) We did not use fused feature since each building sketch in the training set has only one view; 2) The building sketches may have different views than our projections;
3) The building sketches contain richer lines than our projected line drawings.

\section{Conclusions and future work}
\label{sec:future}

We propose what we believe to be the first semi-supervised method for analyzing
and locating decorative style patches on 3D shapes. Our technique utilizes
projective line drawings and multi-view feature encoding for style clustering
and patch extraction. Our semi-supervision is able to take on the same kind of
input, namely, crowd-sourced style ranking triplets, as recent works on style metric
learning~\cite{liu2015style,lun2015elements}. We have shown that comparable
accuracy on style similarity tests can be attained by our method with less
user input than these recent works. As well, we have demonstrated improvements
on style classification and style patch localization over the most recent work
by Hu et al.~\cite{Style17}.

%
%

One of the main limitations of our current style analysis is that by design, it can only extract
stylistic elements that are visually apparent in projected images and localized to the patch level.
These do not include stylistic arrangement of patterns such as those involving symmetries and
repetitions. The main difficulty is that these more global and structural styles may not be fully
visible in projected images.
Technically, our final style patch extraction hinges on the initial pre-selection of representative
feature patches. In addition, for feature-lacking shapes such as a smooth dish or spoon,
projected line drawings cannot be expected to reveal sufficient stylistic elements.
Without sufficient features, our method may lead to undesirable constrained clustering
results.

With the scale of style datasets becoming larger, deep learning will work better on style analysis. However, when the amount of labeled data is small, like in our case, deep learning cannot work well. We have shown that comparable accuracy on style similarity tests can be attained by our method than recent deep learning work Lim et al.~\cite{Lim2016Identifying} on limited style datasets. The main reason why our method works well on relatively small datasets is due to the multi-view feature fusion power of PSLF. On the other hand, our method supports unsupervised and semi-supervised style analysis, in contrast to deep learning which is so far mainly for supervised learning tasks. With our method, we can collect more high-quality labeled data, which can potentially serve for future deep learning research.

%
%
Beyond style-aware shape simplification and view selection, we believe that there is more to
explore on the application front for style patch localization. The ability to identify these
patches spatially allows them to be directly manipulated and applied. For example, style
patches found on the legs of one chair can be transplanted~\cite{takayama2011} to the legs
of another piece of furniture. Also, the style patches can be isolated and encoded as details
over base patches to form a library of style templates. These are both {\em style transfer\/}
tasks. In addition, once we have a set of style patches in possession, we can identify or retrieve
more patches from novel 3D shapes simply based on geometric similarities, either over shape
surfaces or in projective space. Finally, it would be interesting to extend our style analysis
to 3D scenes.

\bibliographystyle{ACM-Reference-Format}
\bibliography{style}


\begin{thebibliography}{00}


\ifx \showCODEN    \undefined \def \showCODEN     #1{\unskip}     \fi
\ifx \showDOI      \undefined \def \showDOI       #1{{\tt DOI:}\penalty0{#1}\ }
  \fi
\ifx \showISBNx    \undefined \def \showISBNx     #1{\unskip}     \fi
\ifx \showISBNxiii \undefined \def \showISBNxiii  #1{\unskip}     \fi
\ifx \showISSN     \undefined \def \showISSN      #1{\unskip}     \fi
\ifx \showLCCN     \undefined \def \showLCCN      #1{\unskip}     \fi
\ifx \shownote     \undefined \def \shownote      #1{#1}          \fi
\ifx \showarticletitle \undefined \def \showarticletitle #1{#1}   \fi
\ifx \showURL      \undefined \def \showURL       #1{#1}          \fi
\providecommand\bibfield[2]{#2}
\providecommand\bibinfo[2]{#2}
\providecommand\natexlab[1]{#1}
\providecommand\showeprint[2][]{arXiv:#2}

\bibitem[\protect\citeauthoryear{Bansal, Shrivastava, Doersch, and
  Gupta}{Bansal et~al\mbox{.}}{2015}]%
        {bansal2015mid}
\bibfield{author}{\bibinfo{person}{Aayush Bansal}, \bibinfo{person}{Abhinav
  Shrivastava}, \bibinfo{person}{Carl Doersch}, {and} \bibinfo{person}{Abhinav
  Gupta}.} \bibinfo{year}{2015}\natexlab{}.
\newblock \showarticletitle{Mid-level Elements for Object Detection}.
\newblock \bibinfo{journal}{{\em arXiv preprint arXiv:1504.07284\/}}
  (\bibinfo{year}{2015}).
\newblock


\bibitem[\protect\citeauthoryear{Chaudhuri, Kakade, Livescu, and
  Sridharan}{Chaudhuri et~al\mbox{.}}{2009}]%
        {chaudhuri2009multi}
\bibfield{author}{\bibinfo{person}{Kamalika Chaudhuri}, \bibinfo{person}{Sham~M
  Kakade}, \bibinfo{person}{Karen Livescu}, {and} \bibinfo{person}{Karthik
  Sridharan}.} \bibinfo{year}{2009}\natexlab{}.
\newblock \showarticletitle{Multi-view clustering via canonical correlation
  analysis}. In \bibinfo{booktitle}{{\em Proc. Int. Conf. on Machine
  Learning}}. \bibinfo{pages}{129--136}.
\newblock


\bibitem[\protect\citeauthoryear{Chen, Tian, Shen, and Ouhyoung}{Chen
  et~al\mbox{.}}{2003}]%
        {chen2003visual}
\bibfield{author}{\bibinfo{person}{D.Y. Chen}, \bibinfo{person}{X.P. Tian},
  \bibinfo{person}{Y.T. Shen}, {and} \bibinfo{person}{M. Ouhyoung}.}
  \bibinfo{year}{2003}\natexlab{}.
\newblock \showarticletitle{On visual similarity based 3{D} model retrieval}.
\newblock \bibinfo{journal}{{\em Computer Graphics Forum\/}}
  \bibinfo{volume}{22}, \bibinfo{number}{3} (\bibinfo{year}{2003}),
  \bibinfo{pages}{223--232}.
\newblock


\bibitem[\protect\citeauthoryear{Chen, Rege, Dong, and Hua}{Chen
  et~al\mbox{.}}{2008}]%
        {chen2008non}
\bibfield{author}{\bibinfo{person}{Yanhua Chen}, \bibinfo{person}{Manjeet
  Rege}, \bibinfo{person}{Ming Dong}, {and} \bibinfo{person}{Jing Hua}.}
  \bibinfo{year}{2008}\natexlab{}.
\newblock \showarticletitle{Non-negative matrix factorization for
  semi-supervised data clustering}.
\newblock \bibinfo{journal}{{\em Knowledge and Information Systems\/}}
  \bibinfo{volume}{17}, \bibinfo{number}{3} (\bibinfo{year}{2008}),
  \bibinfo{pages}{355--379}.
\newblock


\bibitem[\protect\citeauthoryear{Dalal and Triggs}{Dalal and Triggs}{2005}]%
        {dalal2005histograms}
\bibfield{author}{\bibinfo{person}{Navneet Dalal} {and} \bibinfo{person}{Bill
  Triggs}.} \bibinfo{year}{2005}\natexlab{}.
\newblock \showarticletitle{Histograms of oriented gradients for human
  detection}. In \bibinfo{booktitle}{{\em Proc. CVPR}},
  Vol.~\bibinfo{volume}{1}. IEEE, \bibinfo{pages}{886--893}.
\newblock


\bibitem[\protect\citeauthoryear{DeCarlo, Finkelstein, Rusinkiewicz, and
  Santella}{DeCarlo et~al\mbox{.}}{2003}]%
        {decarlo2003suggestive}
\bibfield{author}{\bibinfo{person}{Doug DeCarlo}, \bibinfo{person}{Adam
  Finkelstein}, \bibinfo{person}{Szymon Rusinkiewicz}, {and}
  \bibinfo{person}{Anthony Santella}.} \bibinfo{year}{2003}\natexlab{}.
\newblock \showarticletitle{Suggestive contours for conveying shape}.
\newblock \bibinfo{journal}{{\em ACM Trans. on Graph. (SIGGRAPH)\/}}
  \bibinfo{volume}{22}, \bibinfo{number}{3} (\bibinfo{year}{2003}),
  \bibinfo{pages}{848--855}.
\newblock


\bibitem[\protect\citeauthoryear{Doersch, Gupta, and Efros}{Doersch
  et~al\mbox{.}}{2013}]%
        {doersch2013mid}
\bibfield{author}{\bibinfo{person}{Carl Doersch}, \bibinfo{person}{Abhinav
  Gupta}, {and} \bibinfo{person}{Alexei~A Efros}.}
  \bibinfo{year}{2013}\natexlab{}.
\newblock \showarticletitle{Mid-level visual element discovery as
  discriminative mode seeking}. In \bibinfo{booktitle}{{\em Proc. NIPS}}.
  \bibinfo{pages}{494--502}.
\newblock


\bibitem[\protect\citeauthoryear{Donahue, Jia, Vinyals, Hoffman, Zhang, Tzeng,
  and Darrell}{Donahue et~al\mbox{.}}{2013}]%
        {donahue2013decaf}
\bibfield{author}{\bibinfo{person}{Jeff Donahue}, \bibinfo{person}{Yangqing
  Jia}, \bibinfo{person}{Oriol Vinyals}, \bibinfo{person}{Judy Hoffman},
  \bibinfo{person}{Ning Zhang}, \bibinfo{person}{Eric Tzeng}, {and}
  \bibinfo{person}{Trevor Darrell}.} \bibinfo{year}{2013}\natexlab{}.
\newblock \showarticletitle{Decaf: A deep convolutional activation feature for
  generic visual recognition}.
\newblock \bibinfo{journal}{{\em arXiv preprint arXiv:1310.1531\/}}
  (\bibinfo{year}{2013}).
\newblock


\bibitem[\protect\citeauthoryear{Gal, Sorkine, Mitra, and Cohen-Or}{Gal
  et~al\mbox{.}}{2009}]%
        {gal2009iwires}
\bibfield{author}{\bibinfo{person}{Ran Gal}, \bibinfo{person}{Olga Sorkine},
  \bibinfo{person}{Niloy~J Mitra}, {and} \bibinfo{person}{Daniel Cohen-Or}.}
  \bibinfo{year}{2009}\natexlab{}.
\newblock \showarticletitle{iWIRES: an analyze-and-edit approach to shape
  manipulation}.
\newblock \bibinfo{journal}{{\em ACM Trans. on Graph. (SIGGRAPH)\/}}
  \bibinfo{volume}{28}, \bibinfo{number}{3} (\bibinfo{year}{2009}),
  \bibinfo{pages}{33}.
\newblock


\bibitem[\protect\citeauthoryear{Garces, Agarwala, Gutierrez, and
  Hertzmann}{Garces et~al\mbox{.}}{2014}]%
        {garces2014}
\bibfield{author}{\bibinfo{person}{Elena Garces}, \bibinfo{person}{Aseem
  Agarwala}, \bibinfo{person}{Diego Gutierrez}, {and} \bibinfo{person}{Aaron
  Hertzmann}.} \bibinfo{year}{2014}\natexlab{}.
\newblock \showarticletitle{A Similarity Measure for Illustration Style}.
\newblock \bibinfo{journal}{{\em ACM Trans. on Graph.\/}} \bibinfo{volume}{33},
  \bibinfo{number}{4} (\bibinfo{year}{2014}), \bibinfo{pages}{93:1--9}.
\newblock


\bibitem[\protect\citeauthoryear{Garland and Heckbert}{Garland and
  Heckbert}{1997}]%
        {Garland1997Surface}
\bibfield{author}{\bibinfo{person}{Michael Garland} {and}
  \bibinfo{person}{Paul~S. Heckbert}.} \bibinfo{year}{1997}\natexlab{}.
\newblock \showarticletitle{Surface simplification using quadric error
  metrics}. In \bibinfo{booktitle}{{\em Conference on Computer Graphics and
  Interactive Techniques}}. \bibinfo{pages}{209--216}.
\newblock


\bibitem[\protect\citeauthoryear{Gong, Wang, Guo, and Lazebnik}{Gong
  et~al\mbox{.}}{2014}]%
        {gong2014multi}
\bibfield{author}{\bibinfo{person}{Yunchao Gong}, \bibinfo{person}{Liwei Wang},
  \bibinfo{person}{Ruiqi Guo}, {and} \bibinfo{person}{Svetlana Lazebnik}.}
  \bibinfo{year}{2014}\natexlab{}.
\newblock \showarticletitle{Multi-scale orderless pooling of deep convolutional
  activation features}.
\newblock In \bibinfo{booktitle}{{\em Computer Vision--ECCV 2014}}.
  \bibinfo{publisher}{Springer}, \bibinfo{pages}{392--407}.
\newblock


\bibitem[\protect\citeauthoryear{Hu, Li, Kaick, Huang, Averkiou, Cohen-Or, and
  Zhang}{Hu et~al\mbox{.}}{2017}]%
        {Style17}
\bibfield{author}{\bibinfo{person}{Ruizhen Hu}, \bibinfo{person}{Wenchao Li},
  \bibinfo{person}{Oliver~Van Kaick}, \bibinfo{person}{Hui Huang},
  \bibinfo{person}{Melinos Averkiou}, \bibinfo{person}{Daniel Cohen-Or}, {and}
  \bibinfo{person}{Hao Zhang}.} \bibinfo{year}{2017}\natexlab{}.
\newblock \showarticletitle{Co-Locating Style-Defining Elements on 3D Shapes}.
\newblock \bibinfo{journal}{{\em ACM Transactions on Graphics\/}}
  (\bibinfo{year}{2017}).
\newblock


\bibitem[\protect\citeauthoryear{Huang, Su, and Guibas}{Huang
  et~al\mbox{.}}{2013}]%
        {huang2013fine}
\bibfield{author}{\bibinfo{person}{Qi-Xing Huang}, \bibinfo{person}{Hao Su},
  {and} \bibinfo{person}{Leonidas Guibas}.} \bibinfo{year}{2013}\natexlab{}.
\newblock \showarticletitle{Fine-grained semi-supervised labeling of large
  shape collections}.
\newblock \bibinfo{journal}{{\em ACM Trans. on Graph.\/}} \bibinfo{volume}{32},
  \bibinfo{number}{6} (\bibinfo{year}{2013}), \bibinfo{pages}{190}.
\newblock


\bibitem[\protect\citeauthoryear{Jolliffe}{Jolliffe}{2002}]%
        {jolliffe2002principal}
\bibfield{author}{\bibinfo{person}{Ian Jolliffe}.}
  \bibinfo{year}{2002}\natexlab{}.
\newblock \bibinfo{booktitle}{{\em Principal component analysis}}.
\newblock \bibinfo{publisher}{Wiley Online Library}.
\newblock


\bibitem[\protect\citeauthoryear{Kalogerakis, Chaudhuri, Koller, and
  Koltun}{Kalogerakis et~al\mbox{.}}{2012}]%
        {kalogerakis2012probabilistic}
\bibfield{author}{\bibinfo{person}{Evangelos Kalogerakis},
  \bibinfo{person}{Siddhartha Chaudhuri}, \bibinfo{person}{Daphne Koller},
  {and} \bibinfo{person}{Vladlen Koltun}.} \bibinfo{year}{2012}\natexlab{}.
\newblock \showarticletitle{A probabilistic model for component-based shape
  synthesis}.
\newblock \bibinfo{journal}{{\em ACM Trans. on Graph.\/}} \bibinfo{volume}{31},
  \bibinfo{number}{4} (\bibinfo{year}{2012}), \bibinfo{pages}{55}.
\newblock


\bibitem[\protect\citeauthoryear{Lecun, Bottou, Bengio, and Haffner}{Lecun
  et~al\mbox{.}}{1998}]%
        {Lecun1998Gradient}
\bibfield{author}{\bibinfo{person}{Y. Lecun}, \bibinfo{person}{L. Bottou},
  \bibinfo{person}{Y. Bengio}, {and} \bibinfo{person}{P. Haffner}.}
  \bibinfo{year}{1998}\natexlab{}.
\newblock \showarticletitle{Gradient-based learning applied to document
  recognition}.
\newblock \bibinfo{journal}{{\it Proc. IEEE}} \bibinfo{volume}{86},
  \bibinfo{number}{11} (\bibinfo{year}{1998}), \bibinfo{pages}{2278--2324}.
\newblock


\bibitem[\protect\citeauthoryear{Lee and Seung}{Lee and Seung}{1999}]%
        {lee1999learning}
\bibfield{author}{\bibinfo{person}{Daniel Lee} {and} \bibinfo{person}{Sebastian
  Seung}.} \bibinfo{year}{1999}\natexlab{}.
\newblock \showarticletitle{Learning the parts of objects by non-negative
  matrix factorization}.
\newblock \bibinfo{journal}{{\em Nature\/}} \bibinfo{volume}{401},
  \bibinfo{number}{6755} (\bibinfo{year}{1999}), \bibinfo{pages}{788--791}.
\newblock


\bibitem[\protect\citeauthoryear{Lee, Efros, and Hebert}{Lee
  et~al\mbox{.}}{2013}]%
        {lee2013style}
\bibfield{author}{\bibinfo{person}{Yong~Jae Lee}, \bibinfo{person}{Alexei
  Efros}, {and} \bibinfo{person}{Martial Hebert}.}
  \bibinfo{year}{2013}\natexlab{}.
\newblock \showarticletitle{Style-aware mid-level representation for
  discovering visual connections in space and time}. In
  \bibinfo{booktitle}{{\em Proc. ICCV}}. \bibinfo{pages}{1857--1864}.
\newblock


\bibitem[\protect\citeauthoryear{Li, Zhang, Wang, Cao, Shamir, and Cohen-Or}{Li
  et~al\mbox{.}}{2013}]%
        {li2013curve}
\bibfield{author}{\bibinfo{person}{Honghua Li}, \bibinfo{person}{Hao Zhang},
  \bibinfo{person}{Yanzhen Wang}, \bibinfo{person}{Junjie Cao},
  \bibinfo{person}{Ariel Shamir}, {and} \bibinfo{person}{Daniel Cohen-Or}.}
  \bibinfo{year}{2013}\natexlab{}.
\newblock \showarticletitle{Curve style analysis in a set of shapes}.
\newblock \bibinfo{journal}{{\em Computer Graphics Forum\/}}
  \bibinfo{volume}{32}, \bibinfo{number}{6} (\bibinfo{year}{2013}),
  \bibinfo{pages}{77--88}.
\newblock


\bibitem[\protect\citeauthoryear{Li, Liu, Shen, and Anton}{Li
  et~al\mbox{.}}{2015}]%
        {Li2015Mid}
\bibfield{author}{\bibinfo{person}{Yao Li}, \bibinfo{person}{Lingqiao Liu},
  \bibinfo{person}{Chunhua Shen}, {and} \bibinfo{person}{Van Den~Hengel
  Anton}.} \bibinfo{year}{2015}\natexlab{}.
\newblock \showarticletitle{Mid-level deep pattern mining}. In
  \bibinfo{booktitle}{{\em Proc. CVPR}}. \bibinfo{pages}{971--980}.
\newblock


\bibitem[\protect\citeauthoryear{Lim, Gehre, and Kobbelt}{Lim
  et~al\mbox{.}}{2016a}]%
        {lim2016}
\bibfield{author}{\bibinfo{person}{Isaak Lim}, \bibinfo{person}{Anne Gehre},
  {and} \bibinfo{person}{Leif Kobbelt}.} \bibinfo{year}{2016}\natexlab{a}.
\newblock \showarticletitle{{Identifying Style of 3D Shapes using Deep Metric
  Learning}}.
\newblock \bibinfo{journal}{{\em Computer Graphics Forum\/}}
  \bibinfo{number}{5} (\bibinfo{year}{2016}).
\newblock


\bibitem[\protect\citeauthoryear{Lim, Gehre, and Kobbelt}{Lim
  et~al\mbox{.}}{2016b}]%
        {Lim2016Identifying}
\bibfield{author}{\bibinfo{person}{Isaak Lim}, \bibinfo{person}{Anne Gehre},
  {and} \bibinfo{person}{Leif Kobbelt}.} \bibinfo{year}{2016}\natexlab{b}.
\newblock \showarticletitle{Identifying Style of 3D Shapes using Deep Metric
  Learning}.
\newblock \bibinfo{journal}{{\em Computer Graphics Forum\/}}
  \bibinfo{volume}{35}, \bibinfo{number}{5} (\bibinfo{year}{2016}),
  \bibinfo{pages}{207--215}.
\newblock


\bibitem[\protect\citeauthoryear{Liu and Wu}{Liu and Wu}{2010}]%
        {Liu2010Non}
\bibfield{author}{\bibinfo{person}{Haifeng Liu} {and} \bibinfo{person}{Zhaohui
  Wu}.} \bibinfo{year}{2010}\natexlab{}.
\newblock \showarticletitle{Non-negative matrix factorization with
  constraints}. In \bibinfo{booktitle}{{\em Twenty-Fourth AAAI Conference on
  Artificial Intelligence}}. \bibinfo{pages}{506--511}.
\newblock


\bibitem[\protect\citeauthoryear{Liu, Jiang, Li, Zhou, and Lu}{Liu
  et~al\mbox{.}}{2015b}]%
        {liu2015partially}
\bibfield{author}{\bibinfo{person}{Jing Liu}, \bibinfo{person}{Yu Jiang},
  \bibinfo{person}{Zechao Li}, \bibinfo{person}{Zhi~Hua Zhou}, {and}
  \bibinfo{person}{Hanqing Lu}.} \bibinfo{year}{2015}\natexlab{b}.
\newblock \showarticletitle{Partially Shared Latent Factor Learning With
  Multiview Data.}
\newblock \bibinfo{journal}{{\em IEEE Trans. on Neural Networks \& Learning
  Systems\/}}  \bibinfo{volume}{26} (\bibinfo{year}{2015}),
  \bibinfo{pages}{1233--1246}.
\newblock


\bibitem[\protect\citeauthoryear{Liu, Hertzmann, Li, and Funkhouser}{Liu
  et~al\mbox{.}}{2015a}]%
        {liu2015style}
\bibfield{author}{\bibinfo{person}{Tianqiang Liu}, \bibinfo{person}{Aaron
  Hertzmann}, \bibinfo{person}{Wilmot Li}, {and} \bibinfo{person}{Thomas
  Funkhouser}.} \bibinfo{year}{2015}\natexlab{a}.
\newblock \showarticletitle{Style Compatibility for 3D Furniture Models}.
\newblock \bibinfo{journal}{{\em ACM Trans. on Graph.\/}} \bibinfo{volume}{34},
  \bibinfo{number}{4} (\bibinfo{year}{2015}), \bibinfo{pages}{85:1--85:9}.
\newblock


\bibitem[\protect\citeauthoryear{Lun, Kalogerakis, , Wang, and Sheffer}{Lun
  et~al\mbox{.}}{2016}]%
        {lun2016}
\bibfield{author}{\bibinfo{person}{Zhaoliang Lun}, \bibinfo{person}{Evangelos
  Kalogerakis}, \bibinfo{person}{}, \bibinfo{person}{Rui Wang}, {and}
  \bibinfo{person}{Alla Sheffer}.} \bibinfo{year}{2016}\natexlab{}.
\newblock \showarticletitle{Functionality Preserving Shape Style Transfer}.
\newblock \bibinfo{journal}{{\em ACM Trans. on Graph.\/}} \bibinfo{volume}{35},
  \bibinfo{number}{6} (\bibinfo{year}{2016}), \bibinfo{pages}{209:1--209:14}.
\newblock


\bibitem[\protect\citeauthoryear{Lun, Kalogerakis, and Sheffer}{Lun
  et~al\mbox{.}}{2015}]%
        {lun2015elements}
\bibfield{author}{\bibinfo{person}{Zhaoliang Lun}, \bibinfo{person}{Evangelos
  Kalogerakis}, {and} \bibinfo{person}{Alla Sheffer}.}
  \bibinfo{year}{2015}\natexlab{}.
\newblock \showarticletitle{Elements of style: learning perceptual shape style
  similarity}.
\newblock \bibinfo{journal}{{\em ACM Trans. on Graph.\/}} \bibinfo{volume}{34},
  \bibinfo{number}{4} (\bibinfo{year}{2015}), \bibinfo{pages}{84:1--84:14}.
\newblock


\bibitem[\protect\citeauthoryear{Ma, Huang, Sheffer, Kalogerakis, and Wang}{Ma
  et~al\mbox{.}}{2014}]%
        {ma2014analogy}
\bibfield{author}{\bibinfo{person}{Chongyang Ma}, \bibinfo{person}{Haibin
  Huang}, \bibinfo{person}{Alla Sheffer}, \bibinfo{person}{Evangelos
  Kalogerakis}, {and} \bibinfo{person}{Rui Wang}.}
  \bibinfo{year}{2014}\natexlab{}.
\newblock \showarticletitle{Analogy-driven 3D style transfer}.
\newblock \bibinfo{journal}{{\em Computer Graphics Forum\/}}
  \bibinfo{volume}{33}, \bibinfo{number}{2} (\bibinfo{year}{2014}),
  \bibinfo{pages}{175--184}.
\newblock


\bibitem[\protect\citeauthoryear{Mitra, Wand, Zhang, Cohen-Or, Kim, and
  Huang}{Mitra et~al\mbox{.}}{2013}]%
        {mitra2013}
\bibfield{author}{\bibinfo{person}{Niloy Mitra}, \bibinfo{person}{Michael
  Wand}, \bibinfo{person}{Hao~(Richard) Zhang}, \bibinfo{person}{Daniel
  Cohen-Or}, \bibinfo{person}{Vladimir Kim}, {and} \bibinfo{person}{Qi-Xing
  Huang}.} \bibinfo{year}{2013}\natexlab{}.
\newblock \showarticletitle{Structure-aware Shape Processing}. In
  \bibinfo{booktitle}{{\em SIGGRAPH Asia 2013 Courses}}.
  \bibinfo{pages}{1:1--1:20}.
\newblock


\bibitem[\protect\citeauthoryear{Morley}{Morley}{1999}]%
        {Morley1999The}
\bibfield{author}{\bibinfo{person}{John. Morley}.}
  \bibinfo{year}{1999}\natexlab{}.
\newblock \showarticletitle{The history of furniture : twenty-five centuries of
  style and design in the Western tradition}.
\newblock  (\bibinfo{year}{1999}).
\newblock


\bibitem[\protect\citeauthoryear{Raptis, Kokkinos, and Soatto}{Raptis
  et~al\mbox{.}}{2012}]%
        {raptis2012}
\bibfield{author}{\bibinfo{person}{M. Raptis}, \bibinfo{person}{I. Kokkinos},
  {and} \bibinfo{person}{S. Soatto}.} \bibinfo{year}{2012}\natexlab{}.
\newblock \showarticletitle{Discovering discriminative action parts from
  mid-level video representations}. In \bibinfo{booktitle}{{\em Proc. CVPR}}.
  \bibinfo{pages}{1242--1249}.
\newblock


\bibitem[\protect\citeauthoryear{Razavian, Azizpour, Sullivan, and
  Carlsson}{Razavian et~al\mbox{.}}{2014}]%
        {razavian2014cnn}
\bibfield{author}{\bibinfo{person}{Ali~S Razavian}, \bibinfo{person}{Hossein
  Azizpour}, \bibinfo{person}{Josephine Sullivan}, {and}
  \bibinfo{person}{Stefan Carlsson}.} \bibinfo{year}{2014}\natexlab{}.
\newblock \showarticletitle{CNN features off-the-shelf: an astounding baseline
  for recognition}. In \bibinfo{booktitle}{{\em IEEE CVPR Workshop}}. IEEE,
  \bibinfo{pages}{512--519}.
\newblock


\bibitem[\protect\citeauthoryear{Rusinkiewicz, Cole, DeCarlo, and
  Finkelstein}{Rusinkiewicz et~al\mbox{.}}{2008}]%
        {szymon2008}
\bibfield{author}{\bibinfo{person}{Szymon Rusinkiewicz},
  \bibinfo{person}{Forrester Cole}, \bibinfo{person}{Doug DeCarlo}, {and}
  \bibinfo{person}{Adam Finkelstein}.} \bibinfo{year}{2008}\natexlab{}.
\newblock \showarticletitle{Line Drawings from 3D Models}. In
  \bibinfo{booktitle}{{\em SIGGRAPH Course}}.
\newblock


\bibitem[\protect\citeauthoryear{Sayim and Cavanagh}{Sayim and
  Cavanagh}{2011}]%
        {sayim2011}
\bibfield{author}{\bibinfo{person}{Bilge Sayim} {and} \bibinfo{person}{Patrick
  Cavanagh}.} \bibinfo{year}{2011}\natexlab{}.
\newblock \showarticletitle{What Line Drawings Reveal About the Visual Brain}.
\newblock \bibinfo{journal}{{\em Frontiers in Human Neuroscience\/}}
  \bibinfo{volume}{5} (\bibinfo{year}{2011}), \bibinfo{pages}{118:1--118:4}.
\newblock


\bibitem[\protect\citeauthoryear{Simonyan and Zisserman}{Simonyan and
  Zisserman}{2014}]%
        {Simonyan2014Very}
\bibfield{author}{\bibinfo{person}{Karen Simonyan} {and}
  \bibinfo{person}{Andrew Zisserman}.} \bibinfo{year}{2014}\natexlab{}.
\newblock \showarticletitle{Very Deep Convolutional Networks for Large-Scale
  Image Recognition}.
\newblock \bibinfo{journal}{{\em Computer Science\/}} (\bibinfo{year}{2014}).
\newblock


\bibitem[\protect\citeauthoryear{Singh, Gupta, and Efros}{Singh
  et~al\mbox{.}}{2012}]%
        {singh2012unsupervised}
\bibfield{author}{\bibinfo{person}{Saurabh Singh}, \bibinfo{person}{Abhinav
  Gupta}, {and} \bibinfo{person}{Alexei Efros}.}
  \bibinfo{year}{2012}\natexlab{}.
\newblock \showarticletitle{Unsupervised discovery of mid-level discriminative
  patches}.
\newblock \bibinfo{journal}{{\em Computer Vision--ECCV 2012\/}}
  (\bibinfo{year}{2012}), \bibinfo{pages}{73--86}.
\newblock


\bibitem[\protect\citeauthoryear{Su, Maji, Kalogerakis, and Learned-Miller}{Su
  et~al\mbox{.}}{2015}]%
        {su2015multi}
\bibfield{author}{\bibinfo{person}{Hang Su}, \bibinfo{person}{Subhransu Maji},
  \bibinfo{person}{Evangelos Kalogerakis}, {and} \bibinfo{person}{Erik
  Learned-Miller}.} \bibinfo{year}{2015}\natexlab{}.
\newblock \showarticletitle{Multi-view convolutional neural networks for 3D
  shape recognition}. In \bibinfo{booktitle}{{\em Proceedings of the IEEE
  International Conference on Computer Vision}}. \bibinfo{pages}{945--953}.
\newblock


\bibitem[\protect\citeauthoryear{Takayama, Schmidt, Singh, Igarashi, Boubekeur,
  and Sorkine}{Takayama et~al\mbox{.}}{2011}]%
        {takayama2011}
\bibfield{author}{\bibinfo{person}{Kenshi Takayama}, \bibinfo{person}{Ryan
  Schmidt}, \bibinfo{person}{Karan Singh}, \bibinfo{person}{Takeo Igarashi},
  \bibinfo{person}{Tamy Boubekeur}, {and} \bibinfo{person}{Olga Sorkine}.}
  \bibinfo{year}{2011}\natexlab{}.
\newblock \showarticletitle{{GeoBrush:} Interactive Mesh Geometry Cloning}.
\newblock \bibinfo{journal}{{\em Computer Graphics Forum\/}}
  \bibinfo{volume}{30}, \bibinfo{number}{2} (\bibinfo{year}{2011}),
  \bibinfo{pages}{613--622}.
\newblock


\bibitem[\protect\citeauthoryear{Wang, Ma, Liu, Chen, and Wu}{Wang
  et~al\mbox{.}}{2013b}]%
        {Wang2013Coherence}
\bibfield{author}{\bibinfo{person}{Shan~Dong Wang}, \bibinfo{person}{Zi~Yang
  Ma}, \bibinfo{person}{Xue~Hui Liu}, \bibinfo{person}{Yan~Yun Chen}, {and}
  \bibinfo{person}{En~Hua Wu}.} \bibinfo{year}{2013}\natexlab{b}.
\newblock \showarticletitle{Coherence-enhancing line drawing for color images}.
\newblock \bibinfo{journal}{{\em Science China Information Sciences\/}}
  \bibinfo{volume}{56}, \bibinfo{number}{11} (\bibinfo{year}{2013}),
  \bibinfo{pages}{1--11}.
\newblock


\bibitem[\protect\citeauthoryear{Wang, Gong, Wang, Cohen-Or, Zhang, and
  Chen}{Wang et~al\mbox{.}}{2013a}]%
        {wang2013projective}
\bibfield{author}{\bibinfo{person}{Yunhai Wang}, \bibinfo{person}{Minglun
  Gong}, \bibinfo{person}{Tianhua Wang}, \bibinfo{person}{Daniel Cohen-Or},
  \bibinfo{person}{Hao Zhang}, {and} \bibinfo{person}{Baoquan Chen}.}
  \bibinfo{year}{2013}\natexlab{a}.
\newblock \showarticletitle{Projective analysis for 3D shape segmentation}.
\newblock \bibinfo{journal}{{\em ACM Trans. on Graph.\/}} \bibinfo{volume}{32},
  \bibinfo{number}{6} (\bibinfo{year}{2013}), \bibinfo{pages}{192}.
\newblock


\bibitem[\protect\citeauthoryear{Wikipedia}{Wikipedia}{2016}]%
        {wiki:vis_style}
\bibfield{author}{\bibinfo{person}{Wikipedia}.}
  \bibinfo{year}{2016}\natexlab{}.
\newblock \bibinfo{title}{Style (visual arts) --- Wikipedia{,} The Free
  Encyclopedia}.
\newblock   (\bibinfo{year}{2016}).
\newblock
\showURL{%
\url{https://en.wikipedia.org/w/index.php?title=Style_(visual_arts)&oldid=713614541}}
\newblock
\shownote{[Online; accessed 7-May-2016].}


\bibitem[\protect\citeauthoryear{Xu, Li, Zhang, Cohen-Or, Xiong, and Cheng}{Xu
  et~al\mbox{.}}{2010}]%
        {xu2010style}
\bibfield{author}{\bibinfo{person}{Kai Xu}, \bibinfo{person}{Honghua Li},
  \bibinfo{person}{Hao Zhang}, \bibinfo{person}{Daniel Cohen-Or},
  \bibinfo{person}{Yueshan Xiong}, {and} \bibinfo{person}{Zhi-Quan Cheng}.}
  \bibinfo{year}{2010}\natexlab{}.
\newblock \showarticletitle{Style-content separation by anisotropic part
  scales}.
\newblock \bibinfo{journal}{{\em ACM Trans. on Graph.\/}} \bibinfo{volume}{29},
  \bibinfo{number}{6} (\bibinfo{year}{2010}), \bibinfo{pages}{184:1--184:10}.
\newblock


\bibitem[\protect\citeauthoryear{Xu, Ma, Zhang, Zhu, Shamir, Cohen-Or, and
  Huang}{Xu et~al\mbox{.}}{2014}]%
        {xu2014organizing}
\bibfield{author}{\bibinfo{person}{Kai Xu}, \bibinfo{person}{Rui Ma},
  \bibinfo{person}{Hao Zhang}, \bibinfo{person}{Chenyang Zhu},
  \bibinfo{person}{Ariel Shamir}, \bibinfo{person}{Daniel Cohen-Or}, {and}
  \bibinfo{person}{Hui Huang}.} \bibinfo{year}{2014}\natexlab{}.
\newblock \showarticletitle{Organizing heterogeneous scene collections through
  contextual focal points}.
\newblock \bibinfo{journal}{{\em ACM Trans. on Graph.\/}} \bibinfo{volume}{33},
  \bibinfo{number}{4} (\bibinfo{year}{2014}), \bibinfo{pages}{35}.
\newblock


\bibitem[\protect\citeauthoryear{Zeiler and Fergus}{Zeiler and Fergus}{2014}]%
        {zeiler2014visualizing}
\bibfield{author}{\bibinfo{person}{Matthew~D Zeiler} {and} \bibinfo{person}{Rob
  Fergus}.} \bibinfo{year}{2014}\natexlab{}.
\newblock \showarticletitle{Visualizing and understanding convolutional
  networks}.
\newblock In \bibinfo{booktitle}{{\em Computer Vision--ECCV 2014}}.
  \bibinfo{publisher}{Springer}, \bibinfo{pages}{818--833}.
\newblock


\bibitem[\protect\citeauthoryear{Zelnik-Manor}{Zelnik-Manor}{2004}]%
        {zelnik2004self}
\bibfield{author}{\bibinfo{person}{L. Zelnik-Manor}.}
  \bibinfo{year}{2004}\natexlab{}.
\newblock \showarticletitle{Self-tuning spectral clustering}.
\newblock \bibinfo{journal}{{\em Advances in Neural Information Processing
  Systems\/}}  \bibinfo{volume}{17} (\bibinfo{year}{2004}),
  \bibinfo{pages}{1601--1608}.
\newblock


\bibitem[\protect\citeauthoryear{Zhao, Xu, Zhu, Liu, Zhu, and Yin}{Zhao
  et~al\mbox{.}}{2018}]%
        {zhao2018triangle}
\bibfield{author}{\bibinfo{person}{Yawei Zhao}, \bibinfo{person}{Kai Xu},
  \bibinfo{person}{En Zhu}, \bibinfo{person}{Xinwang Liu},
  \bibinfo{person}{Xinzhong Zhu}, {and} \bibinfo{person}{Jianping Yin}.}
  \bibinfo{year}{2018}\natexlab{}.
\newblock \showarticletitle{Triangle lasso for simultaneous clustering and
  optimization in graph datasets}.
\newblock \bibinfo{journal}{{\em IEEE Transactions on Knowledge and Data
  Engineering\/}} \bibinfo{volume}{31}, \bibinfo{number}{8}
  (\bibinfo{year}{2018}), \bibinfo{pages}{1610--1623}.
\newblock


\bibitem[\protect\citeauthoryear{Zhu, Xu, Chaudhuri, Yi, and Zhang}{Zhu
  et~al\mbox{.}}{2018}]%
        {zhu2018scores}
\bibfield{author}{\bibinfo{person}{Chenyang Zhu}, \bibinfo{person}{Kai Xu},
  \bibinfo{person}{Siddhartha Chaudhuri}, \bibinfo{person}{Renjiao Yi}, {and}
  \bibinfo{person}{Hao Zhang}.} \bibinfo{year}{2018}\natexlab{}.
\newblock \showarticletitle{SCORES: Shape composition with recursive
  substructure priors}.
\newblock \bibinfo{journal}{{\em ACM Transactions on Graphics (TOG)\/}}
  \bibinfo{volume}{37}, \bibinfo{number}{6} (\bibinfo{year}{2018}),
  \bibinfo{pages}{1--14}.
\newblock


\end{thebibliography}

\end{document}